\documentclass[%
prb,
twocolumn,
amsmath,
amssymb,
reprint,
superscriptaddress,
footinbib,
amsmath,amssymb,
aps,
]{revtex4}

\usepackage{graphicx}
\usepackage{dcolumn}
\usepackage{bm}
\usepackage{natbib}
\usepackage{color}
\usepackage{soul}
\usepackage{upgreek}
\usepackage[final]{changes}
\usepackage{color}
\usepackage{hyperref}
\hypersetup{
	colorlinks,
	allcolors=blue,
	linktoc=all
}
\definechangesauthor[name=MA, color=purple]{Miles}

\newcommand*\SiN{Si$_3$N$_4$ }

\begin{document}
\preprint{APS/123-QED}
\title{Photonic chip-based resonant supercontinuum }

\author{Miles H. Anderson}
\affiliation{{\'E}cole Polytechnique F{\'e}d{\'e}rale de Lausanne (EPFL), CH-1015 Lausanne, Switzerland}

\author{Romain Bouchand}
\affiliation{{\'E}cole Polytechnique F{\'e}d{\'e}rale de Lausanne (EPFL), CH-1015 Lausanne, Switzerland}

\author{Junqiu Liu}
\affiliation{{\'E}cole Polytechnique F{\'e}d{\'e}rale de Lausanne (EPFL), CH-1015 Lausanne, Switzerland}

\author{Wenle Weng}
\affiliation{{\'E}cole Polytechnique F{\'e}d{\'e}rale de Lausanne (EPFL), CH-1015 Lausanne, Switzerland}

\author{\\Ewelina Obrzud}
\affiliation{Swiss Center for Electronics and Microtechnology (CSEM), Time and frequency, CH-2002 Neuch\^{a}tel, Switzerland}
\affiliation{Geneva Observatory, University of Geneva, CH-1290 Versoix, Switzerland.}

\author{Tobias Herr}
\affiliation{Swiss Center for Electronics and Microtechnology (CSEM), Time and frequency, CH-2002 Neuch\^{a}tel, Switzerland}

\author{Tobias J. Kippenberg}
\email{tobias.kippenberg@epfl.ch}
\affiliation{{\'E}cole Polytechnique F{\'e}d{\'e}rale de Lausanne (EPFL), CH-1015 Lausanne, Switzerland}




\pacs{Valid PACS appear here}
\maketitle

\textbf{\added{Supercontinuum generation in optical fibers is one of the most dramatic nonlinear effects discovered \cite{alfano_observation_1970, ranka_visible_2000, birks_supercontinuum_2000}, enabling the self-referencing of optical frequency combs and establishing the RF-to-optical link \cite{jones_carrier-envelope_2000, udem_optical_2002} via the formation of multi-octave spanning coherent spectra.}
\added{However, generating coherent supercontinua requires ultrashort pulsed-laser sources with a kilowatt or more in peak-power \cite{nakazawa_coherence_1998,dudley_supercontinuum_2006}, a requirement that becomes harder to maintain as the repetition rate is increased. This has} 
hindered supercontinua at microwave line spacing, i.e. 10s of GHz, ideally suited for optical frequency division\cite{xie_photonic_2017}, Raman spectral imaging\cite{ideguchi_coherent_2013}, telecommunications \cite{marin-palomo_microresonator-based_2017}, or astro-spectrometer calibration \cite{murphy_high-precision_2007}.
\added{Soliton microcombs \cite{herr_temporal_2014, kippenberg_dissipative_2018} by contrast, provide octave-spanning spectra\cite{li_stably_2017, pfeiffer_octave-spanning_2017}, but with good conversion efficiency only at vastly higher repetition rates close to 1 THz.}
Here, we bridge this efficiency gap with resonant supercontinuum, 
\added{requiring pulses with peak powers on the order of single watts, and duration of 1 picosecond. }
By applying synchronous pulse-driving \cite{obrzud_temporal_2017} to a dispersion-engineered, low-loss \SiN photonic chip microresonator\cite{liu_ultralow-power_2018-1}, we generate dissipative Kerr solitons with a strong dispersive wave, both bound to the input pulse.
This creates a smooth, flattened 2,200 line frequency comb, with an electronically detectable repetition rate of 28 GHz, constituting the largest bandwidth-line-count product for any microcomb generated to date.
Strikingly, we observe that solitons exist in a weakly bound state with the input pulse, stabilizing their repetition rate\cite{obrzud_temporal_2017,weng_spectral_2019}, but simultaneously allowing noise transfer from one to the other to be suppressed\cite{weng_spectral_2019,brasch_nonlinear_2019} even for offset frequencies 100 times lower than the linear cavity decay rate. We demonstrate that this nonlinear filtering can be enhanced by pulse-driving asynchronously, in order to preserve the coherence of the comb.
Taken together, our work establishes resonant supercontinuum as a promising route to broadband and coherent spectra. 
}

\begin{figure*} [t]
	\includegraphics{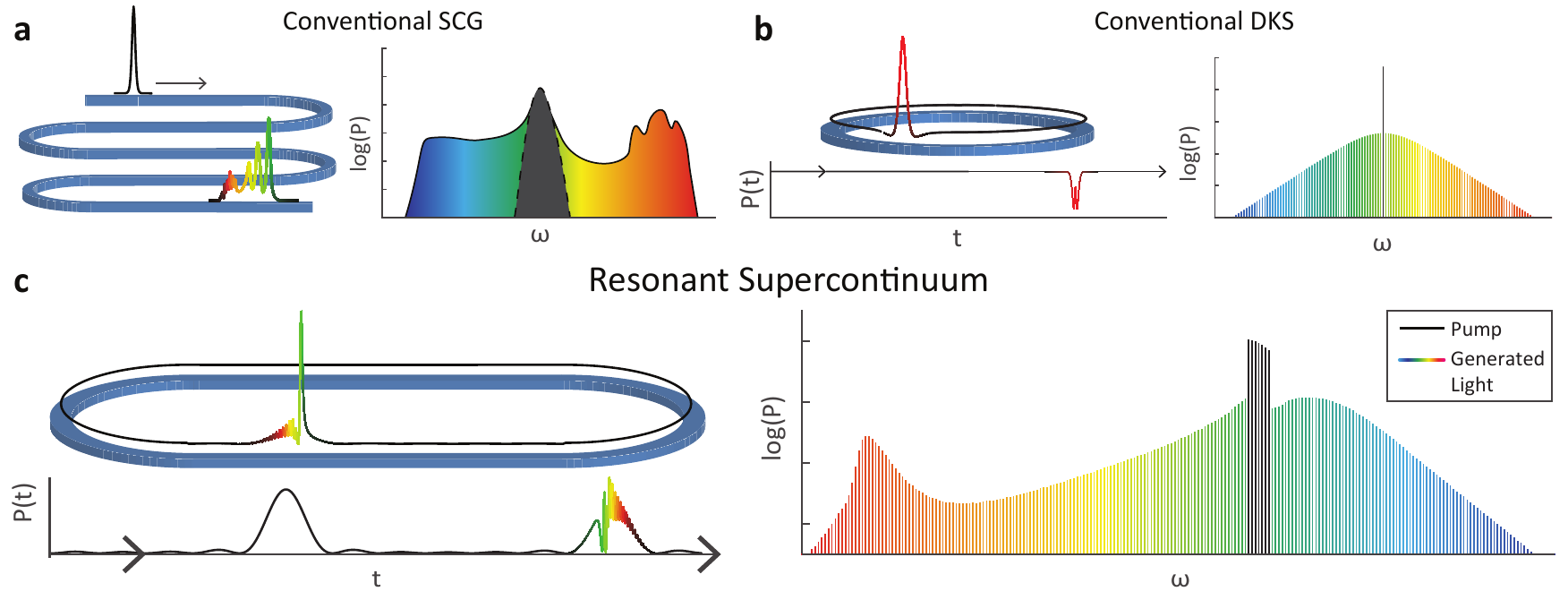}
	\includegraphics{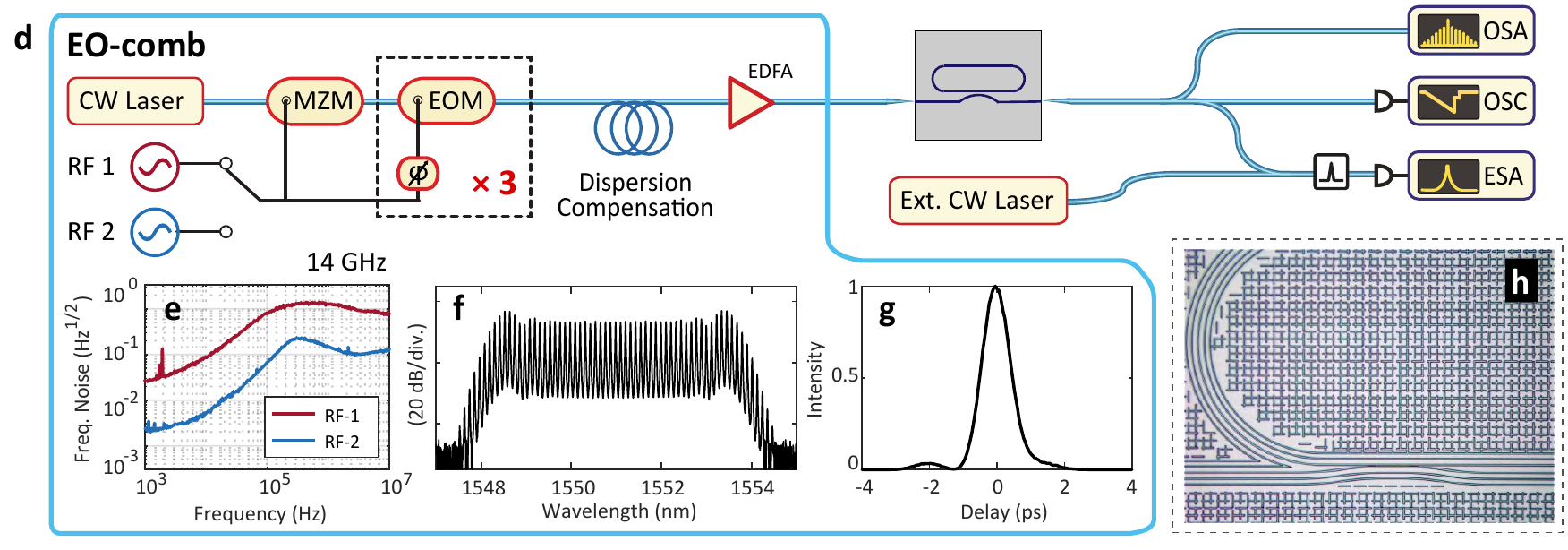}
	\caption{\textbf{Resonant supercontinuum generation using a dispersion engineered \SiN photonic chip }(a-c) Principles of broadband frequency comb generation compared. DKS and Resonant SCG are plotted based on real simulations. (d) Experimental setup. MZM: Mach-Zehnder modulator, EOM: electro-optic modulator, EDFA: erbium-doped fiber amplifier, ESA: electronic spectrum analyzer, OSA: optical spectrum analyzer, OSC: oscilloscope. The input pulse train is coupled into and out of the microresonator chip via lensed fibres. 
	(e) Frequency noise of the two RF signal generators at 14 GHz. (f) Spectrum of the 14 GHz EO-comb before amplification. (g) Retrieved FROG measurement of the optimum pulse duration. (h) Microscope image of part of the microresonator, depicting the coupling section.}
	\label{fig:setup}
\end{figure*}

\begin{figure*} [t]

\centering
\includegraphics{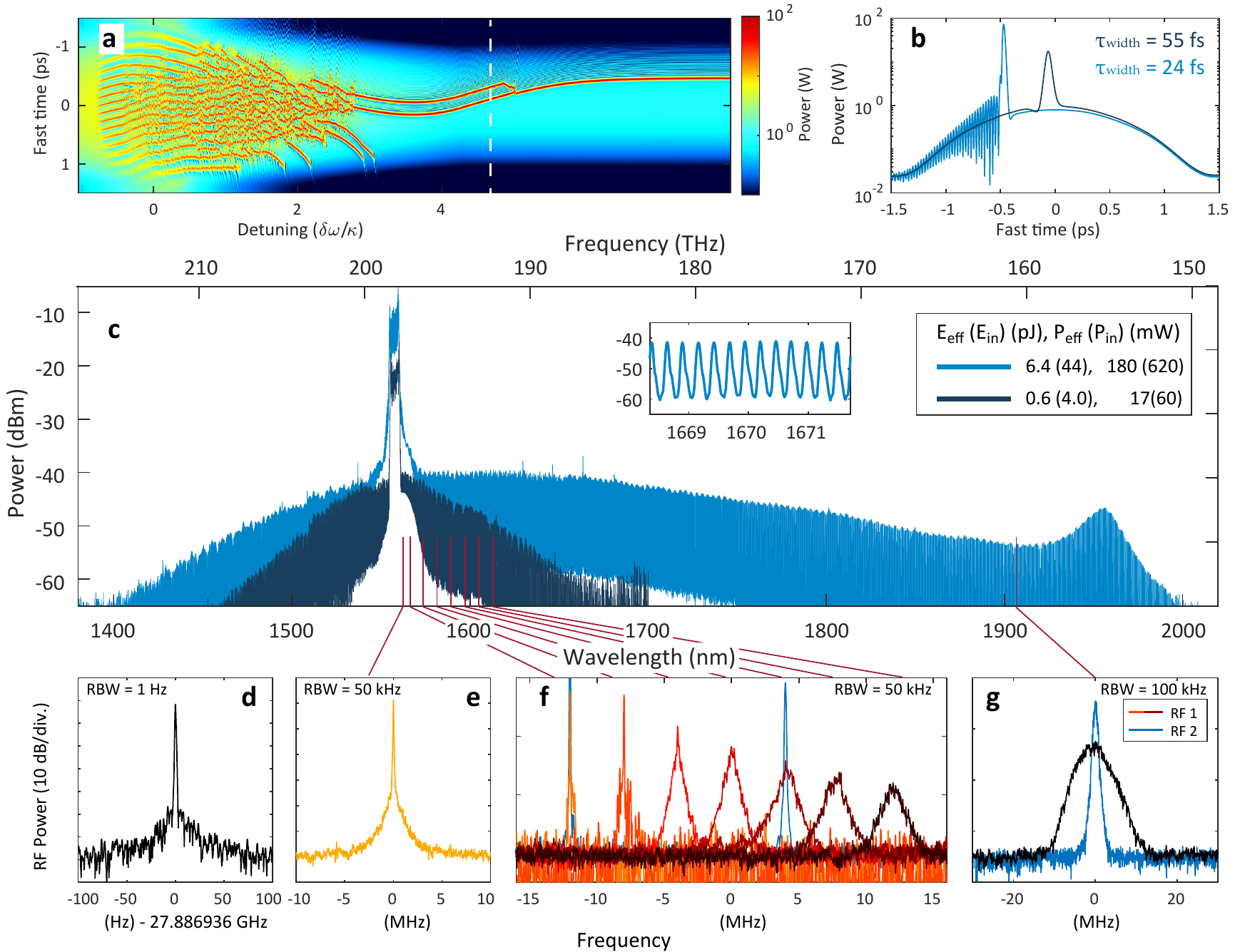}
\caption{\textbf{Resonant supercontinuum generation and coherence. }(a) Simulation of the time-domain intracavity field of the pulse-driven microresonator during a scan over resonance, for a total time of 400$t_\mathrm{photon}$. The scan is halted at $\delta\omega=5\kappa$ (white-dashed line). The pulse background has a repetition rate mismatch $d=-50$ kHz to compensate the offset DKS group velocity. (b) Simulation time-domain slice of the low-power (dark) and high-power (light) dissipative soliton upon the pulse input.
(c) Generated spectrum at maximum detuning, formed using $\sim$1 ps pulses. Lowest energy soliton with minimum pulsed power required in dark color. Fully formed spectrum using higher power in light color. 
Inset: comb lines resolved by the OSA. Input power levels and pulse energies are noted as \emph{effective} power coupled to the microresonator (total power incident on chip). See Methods for calculation. (d) The soliton repetition rate beatnote. 
(e) Beatnote at the 12th line at the edge of the EO-comb. (f) Beatnotes of $\mu$th comb line, measured between 1560--1620 nm, from left-to-right $\mu=$ -45, -88, -132, -175, -217, -259, -300 (horizontally offset by 4 MHz for clarity). 
(g) Beatnote at 1908 nm ($\mu=-1300$).}
\label{fig:results1}

\end{figure*}

Supercontinuum generation (SCG, or `white light' generation \cite{bellini_phase-locked_2000}) is a process where high intensity optical pulses are converted into coherent octave-spanning spectra by propagation through a dispersion-engineered waveguide, fiber, or material (Fig. \ref{fig:setup}(a)). Following the demonstration of dramatic broadening in optical fiber \cite{ranka_visible_2000}, the process has been well studied in photonic crystal fibers \cite{russell_photonic-crystal_2006,dudley_supercontinuum_2006}, owing to their capacity for dispersion engineering. 
SCG is based on a combination of nonlinear phenomenon including soliton fission, dispersive wave formation, and the Raman self-frequency shift \cite{skryabin_colloquium:_2010}. 
Commonly, in order to generate a supercontinuum which is coherent as well as having ultra-high bandwidth, ultrashort pulses ($\sim$100 fs) with high peak powers (1 kW) are needed so that the pulse undergoes a process known as soliton fission \cite{herrmann_experimental_2002, skryabin_colloquium:_2010}, as opposed to incoherent modulation instability\cite{nakazawa_coherence_1998}. Dispersive wave emission (alternatively soliton Cherenkov radiation\cite{akhmediev_cherenkov_1995}) simultaneously serves to extend the spectrum towards other spectral regions far from the pump \cite{hilligsoe_initial_2003}. 
To achieve this, SCG has most often required the input of mode-locked laser systems operating at repetition rates of $<$1 GHz so as to provide \added{high peak power}. Although photonic chip-based waveguides with a high material nonlinearity have reduced required pulse energies by an order of magnitude, and have allowed lithographic dispersion engineering \cite{yeom_low-threshold_2008,halir_ultrabroadband_2012, leo_dispersive_2014, guo_mid-infrared_2018}, synthesis of octave spanning spectra with line spacing \textgreater10 GHz has remained challenging. 
Accessing this regime has been achieved using SCG driven with electro-optic frequency combs \cite{wu_supercontinuum-based_2013,beha_electronic_2017, obrzud_broadband_2018-1, carlson_ultrafast_2018}, providing ultrabroad frequency comb formation at repetition rates of 10-30 GHz, although multiple stages of amplification and pulse-compression were required in order to replicate the same pulse duration and peak powers available from mode-locked lasers.

An alternative technique for the generation of coherent frequency comb spectra is Kerr comb generation \cite{kippenberg_dissipative_2018}, i.e. soliton microcombs. 
Kerr comb generation uses the resonant build-up of a continuous-wave laser to generate a frequency comb via parametric frequency conversion and the formation of dissipative Kerr solitons (DKS) \cite{herr_temporal_2014}. These DKS exhibit a rich landscape of dynamical states, such as breathing \cite{leo_dynamics_2013,lucas_breathing_2017}, chaos \cite{anderson_observations_2016}, and bound-states \cite{wang_universal_2017}. 
In contrast to SCG, DKS circulate indefinitely and are a soliton of an `open system', relying on a double balance of nonlinearity and dispersion, as well as parametric gain and dissipation \cite{akhmediev_dissipative_2008}. 
The cavity enhances the pump field, dramatically reducing the input power threshold for soliton formation. Yet, the process itself has an efficiency that reduces with decreasing repetition rate owing to the reduced overlap of the DKS and the background pump \cite{bao_nonlinear_2014} (Fig. \ref{fig:setup}(b)). 
As a consequence, octave-spanning soliton microcombs to date have been synthesized with 1 THz line spacing\cite{pfeiffer_octave-spanning_2017,li_stably_2017}, and it has proven challenging to synthesize spectra with 10-50 GHz repetition rate with either SCG or microcomb formation. However, a growing number of applications benefit from coherent supercontinua with line spacing in the microwave domain that can be easily detected and processed by electronics. Such widely-spaced comb spectra are resolvable in diffraction-based spectrometers for astrocombs \cite{ewelina_obrzud_microphotonic_2019,suh_searching_2019}, and are highly appropriate as sources for massively parallel wavelength-division multiplexing \cite{marin-palomo_microresonator-based_2017, hu_single-source_2018}. They can also remove the ambiguity in the identification of individual comb lines.

In this work we demonstrate \emph{resonant} supercontinuum generation, a synthesis between conventional SCG and soliton microcombs (Fig. \ref{fig:setup}(c)). By supplying a microresonator with a pulsed input, we take equal advantage of the resonant enhancement offered by the cavity, as well as the higher peak input powers and conversion efficiency allowed by pulses as compared to CW\cite{malinowski_optical_2017}. Where recent works on pulse-driven Kerr cavities for DKS generation have focused on facilitating access to single soliton generation with high conversion efficiency \cite{obrzud_temporal_2017}, and peak-power enhancement \cite{lilienfein_temporal_2019}, the use of dispersion-optimized photonic waveguides to generate a spectrum with an enhanced bandwidth and flatness has not yet been demonstrated with this method.
In our work, we make use of the low-loss photonic \SiN resonator platform \cite{okawachi_bandwidth_2014}. By promoting low dispersion with a strong third-order component, we generate a flattened, broadband spectrum close to 2/3 of an octave wide, \added{using pulses ten time longer in duration, and with peak power 2 orders of magntiude lower, than} in conventional \SiN-based SCG \cite{carlson_ultrafast_2018, okawachi_carrier_2018} and with an electronically detectable repetition rate of 28 GHz.

We further investigate, numerically and experimentally, the nature of the bonding between the generated DKS and the driving pulse\cite{hendry_spontaneous_2018}, particularly the \emph{nonlinear} filtering\cite{brasch_nonlinear_2019} of noise transfer this weak bonding gives rise to. This nonlinear filtering is found, remarkably, to combat noise multiplication, a known drawback of EO-comb-driven SCG, where frequency noise on the input pulse repetition rate is transferred and multiplied over the generated optical lines destroying their coherence \cite{ishizawa_phase-noise_2013, beha_electronic_2017}. We find a way to maximize this filtering, in both simulations and experiment.


\textbf{Resonant Supercontinuum Results.} The chip-based \SiN microresonator used for this experiment (a section depicted in Fig. \ref{fig:setup}(h)), has a free spectral range (FSR) of 27.88 GHz and a loaded linewidth in the telecom band of \added{$\kappa=2\pi.$110 MHz} (most probable value\cite{liu_ultralow-power_2018-1}). The waveguide dimensions have been selected to give a low dispersion of $\beta_2=-11$ fs$^2/$mm. The pulse-train incident on this chip is synthesized using cascaded electro-optic modulation, intensity modulation, and dispersion compensation \cite{kobayashi_optical_1988,fujiwara_optical_2003} (see Fig. \ref{fig:setup}(d)), providing pulses with a minimum duration of 1 ps, at a repetition rate $f_\mathrm{eo}=$ 13.94 GHz. In this way, the microresonator is sub-harmonically pumped every two roundtrips\cite{ewelina_obrzud_microphotonic_2019}. 
This decreases the conversion efficiency by a factor of 2, but reduces the requirements on the microwave transmission system. A tunable RF signal generator supplies $f_\mathrm{eo}$, and we keep two alternative RF sources -- with relatively high (RF-1) and low (RF-2) phase-noise respectively -- in order to observe how their frequency noise is transferred to the resonant supercontinuum. Further details are given in Methods.

DKS states are generated on the input pulses by sweeping their carrier frequency $\omega_p$ from the blue- to the red-detuned side of the cavity resonance $\omega_0$, to the region of cavity bistability \cite{herr_temporal_2014}, such that the detuning $\delta\omega=\omega_0-\omega_p>0$. Before a DKS can be formed stably, the difference between the repetition rate of the pulse-train $f_\mathrm{eo}$ and the cavity FSR has to be matched to within a `locking range'. 
Inside this range, the generated soliton becomes locked to the driving pulse \cite{obrzud_temporal_2017} (or modulated background \cite{weng_spectral_2019}), so that the comb $f_\mathrm{rep}=f_\mathrm{eo}$. A simulated example of this is depicted in Fig. \ref{fig:results1}(a). In this experiment, the locking range is $\sim$30\textendash50 kHz.

\begin{figure*} [t]

\includegraphics[width=\textwidth]{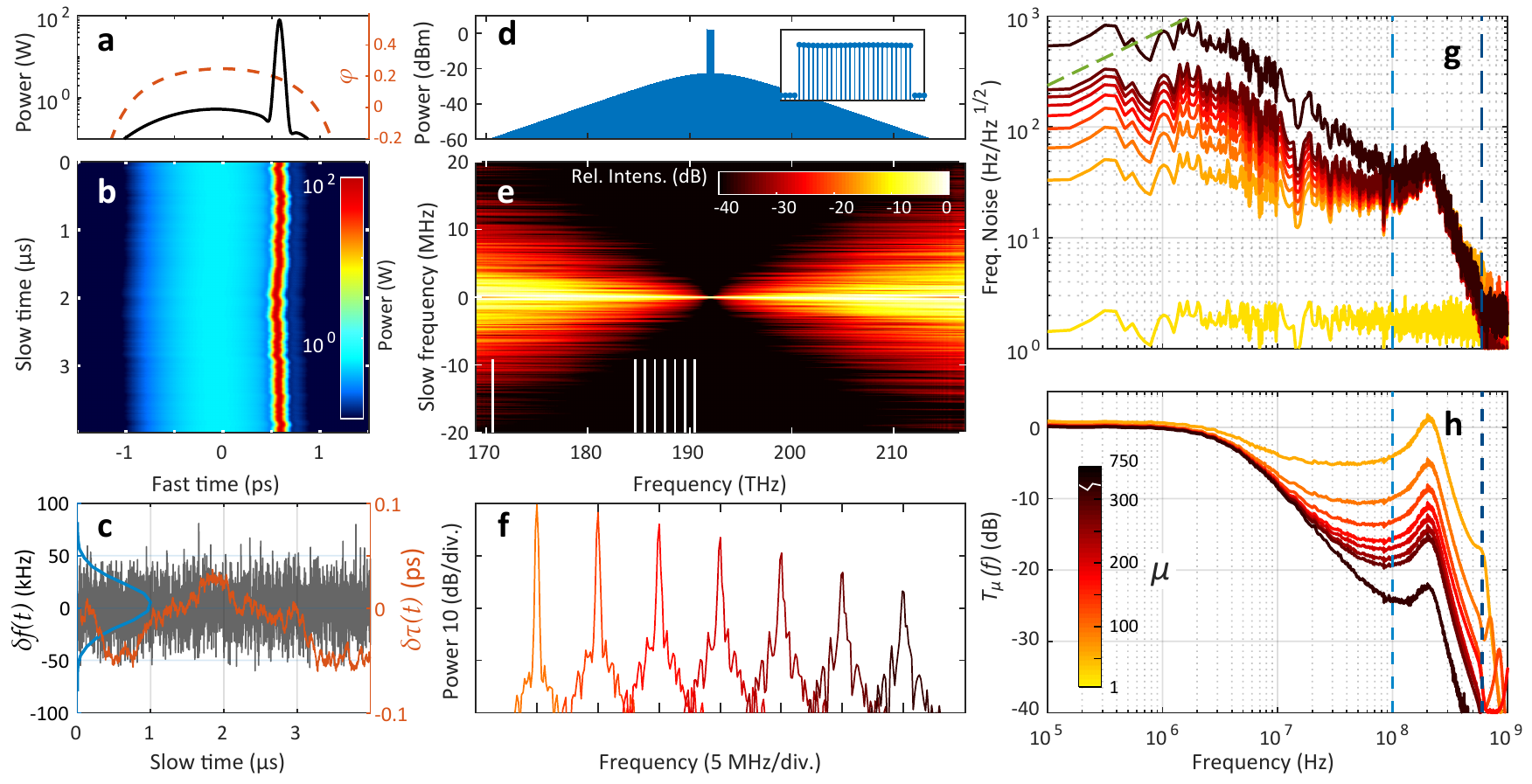}
\caption{\textbf{Simulation of noise multiplication and nonlinear filtering, based on the Lugiato-Lefever Equation. }(a) Intracavity field at time $t$. Instantaneous phase of the background pulse as dotted line. (b) Slow time vs. fast time graph of intracavity field. (c) Frequency noise (gray, histogram in blue), and corresponding absolute timing jitter (orange), imposed on the input pulse over the period of the simulation. (d) Output spectrum, with inset showing input spectrum. (e) Slow frequency vs. fast frequency graph of the long-term frequency comb, with columns normalized to peak. White marks indicate comb lines depicted in (g) and (h). (f) Individual noisy comb lines selected from (d), with same comb mode $\mu$ as the experiment Fig. \ref{fig:results1}(f). (g) Frequency noise spectra and (h) corresponding RF-noise transfer functions for individual comb lines marked in (e), including a comb line far from center ($\mu=-750$). Dotted lines indicate the linear cavity bandwidth (light blue), cavity detuning (dark blue), and $\beta$-line (green). Traces are averaged from 8 simulations for clarity.}
\label{fig:simulation1}
\end{figure*}

The measured output spectrum of the microresonator during single-state DKS operation are presented in Fig. \ref{fig:results1}(c), generated at two input powers: the minimum required to form a DKS, and a higher power generating the most energetic spectrum for this work. They both exist at the maximum accessible cavity detuning $\delta\omega$, where the spectral bandwidth of the DKS scales as $\Omega_{S}\propto\sqrt{P_0}$ \cite{coen_universal_2013,lucas_detuning-dependent_2017}, hence the dramatic broadening of the spectra. The first, least energetic soliton has a 3-dB bandwidth of 9.5 THz, and an estimated pulse duration of 55 fs based on a sech$^2$ fit. The energy of a soliton scales the same way \cite{bao_nonlinear_2014}, hence this first DKS has the highest conversion efficiency from input comb to generated lines of 8\%. The high-energy DKS measurably spans 64 THz or 600 nm, accounting for 2,300 measurable lines (1,400 in 10 dB), and has a conversion efficiency of 2.8\%. This is the highest line-count for a single-state DKS, with a bandwidth exceeding the C+L bands, to our knowledge. Simulations shown in Fig. \ref{fig:results1}(a,b) replicating the measured spectrum predict a distorted DKS due to the strong dispersive wave emission, and having a duration of $\sim24$ fs.
The spectrum is strongly enhanced on the long-wavelength side due to the prominent third-order dispersion of the waveguide, forming a dispersive wave at 1957 nm, combined with the soliton Raman self-frequency shift which has shifted the spectral center towards 1590 nm \cite{karpov_raman_2016, yi_theory_2016}. 

Importantly, no fast-tuning methods\cite{lamb_optical-frequency_2018} were required in order to form DKS. 
Piezo-tuning was sufficient, suggesting a practical absence of cavity thermal relaxation, which has complicated stable soliton generation in the past \cite{brasch_bringing_2016, li_stably_2017}. The number of DKS ranged from 1 to 3.
The effective average power (see Methods) required to generate these single-soliton states ranges from 18 to 180 mW, which is highly efficient considering the $Q$ of the resonator as compared to recent experimental work in CW-driven \SiN microresonators, of similar FSR, with even higher $Q$ \cite{liu_ultralow-power_2018-1, liu_nanophotonic_2019}.
The corresponding pulse energies range from 0.6 to 6 pJ, and we estimate the peak powers to be from \added{0.6 to 6 W based on the FROG measurement of the pulse profile}. Details on the simulation, multi-state spectra, EO-comb pulse compression can be found in the S.I.

Looking closely at the central EO-comb \added{pump spectrum (Fig. \ref{fig:results1}(c) around 1560 nm)}, one can see the minute amount of broadening due to self-phase modulation of the pulse as it travels through the 5 mm of \SiN waveguide, representing the equivalent `conventional supercontinuum' occurring on a 1 ps, sub-10 pJ pulse. Comparing this to the broadband DKS spectrum, produced by the same pulse incident on a resonator of the \emph{same length} of \SiN waveguide -- 5 mm -- this puts the effect of resonant supercontinuum into stark contrast with its conventional counterpart. 

\textbf{Coherence properties. }Fig. \ref{fig:results1}(d) shows the repetition rate beatnote of the DKS \emph{excluding} the EO-comb spectrum. The beatnote corresponds exactly to 2 times the RF source frequency, and with a 1 Hz limited linewidth demonstrates high repetition rate stability. For measuring the optical coherence of the comb, optical heterodyne measurements are taken against increasing values of $\mu$, the comb line index from the center pump, from 1550 nm to the outer edge at 1908 nm plotted in Fig. \ref{fig:results1}(e-g). Fig \ref{fig:results1}(e) shows a narrow heterodyne beatnote at the edge of the EO-comb. 
As comb lines become further away from the center, their linewidth quickly broadens as shown in Fig. \ref{fig:results1}(f), where we plot heterodyne beatnotes up to a range of 70 nm from the comb center. This noise multiplication continues to the long-wavelength edge of the comb, where the heterodyne beatnote with a narrow-linewidth 1908 nm laser is plotted in Fig. \ref{fig:results1}(g). Here, the linewidth has expanded to around 7.5 MHz according to Gaussian fitting. 

When we switch our RF signal generator from RF-1 to the lower noise source RF-2, and measure the heterodyne comb beatnotes at the same wavelengths, three examples of which are plotted 
in Fig. \ref{fig:results1}(f,g), we find that linewidths are decisively more narrow. The linewidth at 1908 nm in particular has reduced by almost a factor of 10, down to 900 kHz. This difference in coherence at the wings of the spectrum, using different EO-comb RF sources, confirms to us that this is the result of RF noise multiplication, imposed on the comb spacing through the locking between the input pulse and the DKS.
However, as shown in the S.I., when we form a conventional supercontinuum using an EO-comb driven by the same RF sources, we can detect the beatnote at 1908 nm \emph{only} when RF-2 is used, not RF-1. This indicates the presence of an additional filtering effect possesd only by the DKS. 


\begin{figure*} [t]
	
	\includegraphics[width=\textwidth]{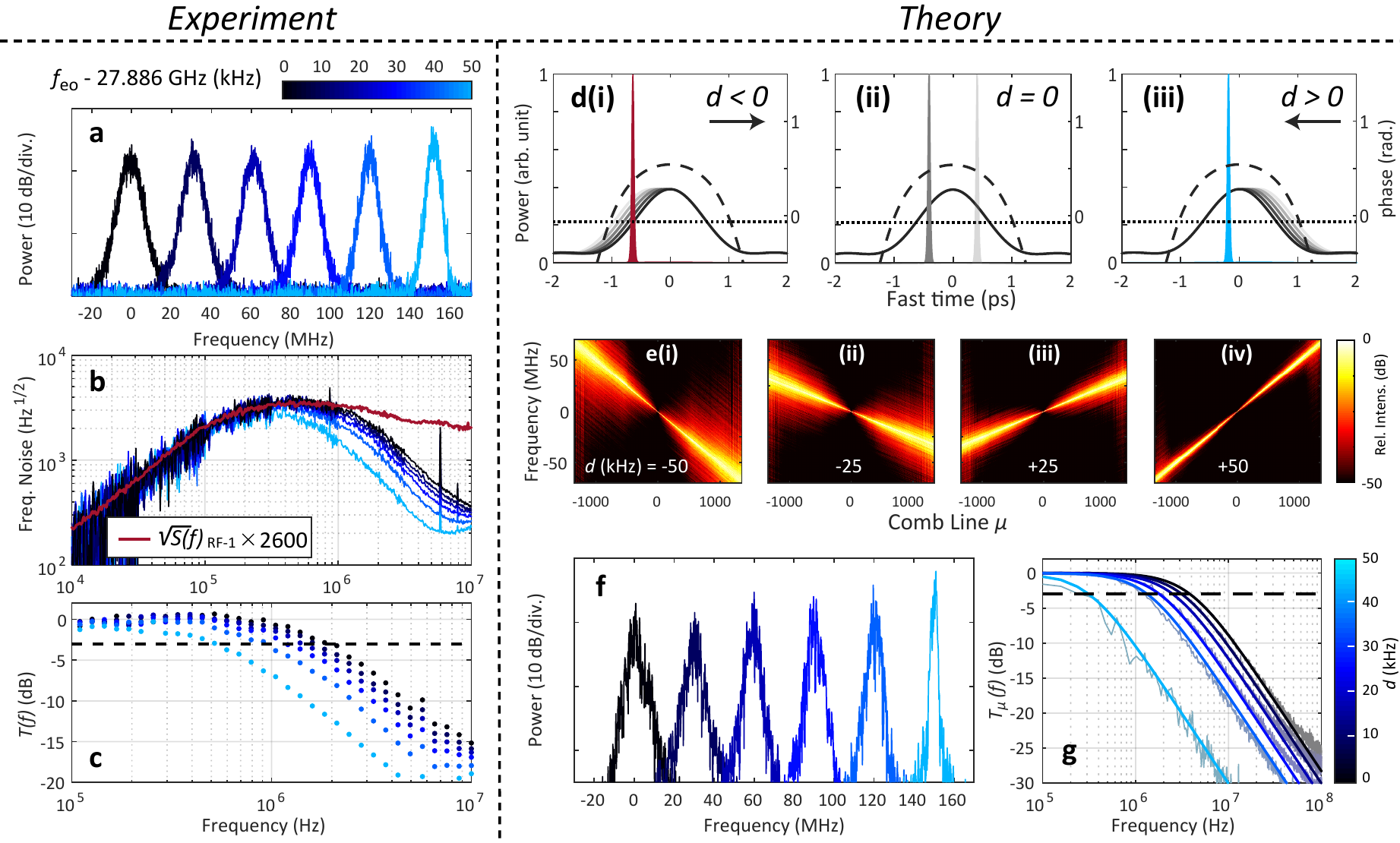}
	\caption{\textbf{Optimization of soliton-based nonlinear filtering via asynchronous driving: experiment and simulation}. (a) Experimental DKS heterodyne beatnote at 1908 nm ($\mu=-1300$) as $f_\mathrm{eo}$ is varied over a range of 50 kHz. Beatnotes horizontally offset each by 30 MHz for clarity. (b) Corresponding measured frequency noise, plotted with the multiplied frequency noise of RF-1. (c) Experimental transfer function based on (b), with 3-dB level marked with dashed line.
	(d)i-iii Conceptual soliton trapping locations on a chirped pulse background under different values of de-synchronized driving $d$. Arrow indicates background repetition-rate mismatch, with pulse phase profile (dashed) and $F_C$ (dotted). (e)i-iv Simulated slow vs. fast frequency graphs for different values of de-synchronization $d$. (f) Slice of the $\mu=-1300$ beatnote (1908 nm) in slow frequency for $d$ from 0 to 50 kHz (blue). (g) Simulated noise transfer functions for $d$ from 0 to 50 kHz. Low-pass fit profile in colored line, original results in gray.
	}
	\label{fig:simulation2}
	
\end{figure*}


\textbf{Noise Transfer Simulation.} If one were to assume that the generated soliton is \emph{perfectly} locked to the input pulse, we would expect the optical frequency noise of each soliton comb line to be coupled to the RF noise on the input pulse repetition rate $f_\mathrm{eo}$, such that $S^{(\mu)}_{\delta\nu}(f)=\mu^2S_f^{\mathrm{(rf)}}(f)$, with relative comb index $\mu$ (assuming other sources of laser noise are small by comparison). However this is not the case for a DKS.
The frequency-noise multiplication transfer function 
can be found in numerical simulations based on the Lugiato-Lefever Equation (LLE) \cite{lugiato_spatial_1987} with parameters similar to that of a typical \SiN resonator,  where instead of a normally CW-driving term we use a pulsed input $F(\phi)$ (where $\phi$ is the spacial coordinate of the cavity) similar in duration to that used in the experiment. It is also slightly positively chirped, in line with our experimental EO-comb compression stage (see Methods and the S.I.), giving a negative phase curvature on the pulse \cite{AGRAWAL199560}. Frequency noise equivalent to a uniform power-spectral density of $S_f^{\mathrm{(rf)}}(f)=$ 1.0 Hz$^2$/Hz is applied to the input pulse over a long period of `slow time' ($t>2,000t_\mathrm{photon}$), and the corresponding jitter of the soliton is captured. For this first simulation, we have set $f_\mathrm{eo}$, the input pulse repetition rate, to be equal to the FSR for fully synchronous driving.

The simulation results are shown in Fig. \ref{fig:simulation1}. In the time domain (shown in Fig. \ref{fig:simulation1}(a,b)), the generated soliton is located at its `trapping point' at $\sim0.6$ ps at the trailing edge of the pulse\cite{hendry_spontaneous_2018}, but under these synchronous and symmetrical pulse conditions, it may equally find itself at -0.6 ps on the leading edge. As it's trapped, or locked, to the background input pulse, it inherits jitter and gradually walks back and forth in the `fast time' domain. The noise of the input pulse itself, and its corresponding walk in the time domain, is plotted in Fig. \ref{fig:simulation1}(c).

The corresponding frequency domain results are shown in Fig. \ref{fig:simulation1}(d,e). Fig. \ref{fig:simulation1}(e) in particular is obtained by taking the Fourier transform over both dimensions of the optical field in Fig. \ref{fig:simulation1}(b), therefore plotting the power spectral densities on the y-axis of each individual comb line along the x-axis -- a simulated heterodyne beatnote (normalized to peak). 
As is evident, the linewidth of each comb line widens considerably as they become further from the comb center. 
In Fig. \ref{fig:simulation1}(f), individual beatnotes corresponding to the same comb line measurements shown in Fig. \ref{fig:results1}(f) show good qualitative agreement.

The profile of this frequency noise transfer to individual comb lines is given in Fig. \ref{fig:simulation1}(g,h), where we plot their frequency noise spectra and corresponding normalized transfer functions $T^{(\mu)}(f)=S^{(\mu)}_{\delta\nu}/(\mu^2S^\mathrm{(rf)}_f)$ respectively. As expected for low offset frequencies, noise power is fully multiplied by $\mu^2$, fulfilling the requirement for the soliton to be locked to the input pulse over the long term.
Strikingly however, above some cut-off frequency $f_c\approx$ 3 MHz, the transfer of noise power drops significantly at a slope of -20 dB/decade, showing how the soliton is able to `ignore' fast background motion of the input pulse despite being locked to it over the long-term. Interestingly, this cut-off frequency is on the order of 100 times lower than the linear cavity bandwidth for this simulated \SiN cavity, of 100 MHz, demonstrating that this filtering is born of the nonlinear DKS regime, as has been observed in MgF$_2$ crystalline microresonators\cite{weng_spectral_2019}. 
Beyond this cut-off point, the transfer function begins to be dominated by the response of the cavity \cite{guo_universal_2017}, where we see a resonance located after the cavity bandwidth at $\kappa$, and a further strong cut-off at the cavity detuning at $\delta\omega=2\pi\cdot600$ MHz. The exact nature of these resonances is beyond the scope of this work, though the simulated traces presented here are in excellent qualitative agreement with the numerical and experimental results presented in an independent and concurrent work by \emph{Brasch et al.}\cite{brasch_nonlinear_2019}, where the full linear and nonlinear response is investigated in detail. 
Taking the $\beta$-line as a guide, these far-offset features will not factor into the linewidth of the outer comb lines \cite{domenico_simple_2010} in this system.

\textbf{Optimization of Nonlinear Filtering. }Returning attention to the experimental heterodyne beatnote measurement at 1908 nm, we have observed an interesting effect when the driving repetition $f_\mathrm{eo}$ is varied. Fig \ref{fig:simulation2}(a) shows the 1908 nm beatnote as $f_\mathrm{eo}$ is swept from the minimum to the maximum of the soliton locking range (0 kHz defined as the minimum). The linewidth appears to narrow, reaching a minimum at the upper edge of the locking range, in this case 50 kHz. 
To characterize this narrowing phenomenon, we measured the frequency noise spectrum of this beatnote using in-phase/quadrature analysis \cite{schiemangk_accurate_2014} as $f_\mathrm{eo}$ is varied across the locking range, which is shown in Fig. \ref{fig:simulation2}(b). Also overlaid is $\mu\sqrt{S^\mathrm{(rf1)}_f}$, where $S^\mathrm{(rf1)}_f$ here is the independently measured frequency noise spectrum of the signal generator RF-1 at 14 GHz, with comb line $\mu=2\times1300$, (factor of 2 being from the half rep-rate driving). 
We further plot the corresponding experimental transfer functions $T^{(\mu)}(f)$, as before, in Fig. \ref{fig:simulation2}(c). As shown, the frequency noise level of the 1908 nm beatnote very closely follows the multiplied RF noise until a certain cut-off frequency, which varies from a maximum of 2 MHz reducing to $\sim$500 kHz at the edge of the locking range. In excellent qualitative agreement with the simulation results in Fig \ref{fig:simulation1}, the value of this corner frequency is on the order of 100 times less than the linear cavity bandwidth, of 110 MHz, experimentally confirming the presence of nonlinear filtering.


We next carry out numerical simulations to analyze the $f_\mathrm{eo}$-dependent nonlinear filtering behavior. We apply a mismatch between the input pulse train repetition rate and the native repetition rate of the soliton ($d=f_\mathrm{eo}-F\!S\!R$). A small level of positive chirp on the input pulse is included in the simulation as per experimental condition (see Methods). Fig. \ref{fig:simulation2}(e) show the same type of result as Fig. \ref{fig:simulation1}(e), only now for a DKS comb for 4 different values of $d$ (manifesting as the gradient in the comb line centers) between -50 and 50 kHz. The effect on the comb linewidth broadening is dramatic. The simulated comb line for $\mu=-1300$ as $d$ is varied between 0 and 50 kHz (the maximum of the locking range) is displayed in Fig. \ref{fig:simulation2}(f), showing excellent agreement with the experimental observation in Fig. \ref{fig:simulation2}(a).

Our experimental and simulation results reveal that as the repetition rate mismatch $d$ changes, the `trapping' location of the DKS on the driving pulse, as well as the local trapping gradient can be significantly different. Previous studies have demonstrated that solitons can acquire a non-zero drift across the cavity space due to the presence of a gradient on the background driving field. Specifically, for a purely phase-modulated background, the soliton will become attracted to the \emph{peak} of the phase profile \cite{jang_temporal_2015}. Conversely, for pure amplitude-variation, a soliton will become attracted to the \emph{edge} of the pulse, at some critical intensity level $F_C$ \cite{hendry_spontaneous_2018, hendry_impact_2019-1}. In our experiment, the driving field is essentially a mixture of both amplitude modulation (pulse driving) and phase modulation (additional chirping). As a result, the soliton is drawn towards an \emph{intermediate} trapping location between the intensity-based trap at the edge of the pulse, and the phase-based trap at the peak. This trapping point will be modified by $d$, which acts as an effective force \cite{javaloyes_dynamics_2016}. In order for a DKS to continue to sustain itself, it must follow the pulse at its own shifted repetition rate, so that:
\begin{equation}
2\pi d + \frac{\partial\phi_S}{\partial t} = 0
\end{equation}
where $\phi_S$ is the angular coordinate of the soliton inside the resonator. If $d$ is non-zero, the soliton must move to a location in order to acquire a shifted repetition rate due the gradient in background phase and/or intensity\cite{jang_temporal_2015,hendry_impact_2019-1}. Fig. \ref{fig:simulation2}(d) illustrates these situations from (i) to (iii), where $d<0$, $d=0$, and $d>0$. The soliton is initially positioned on the left, leading edge of the pulse. In Fig. \ref{fig:simulation2}d(i), the mismatch $d$ and the intensity-based trapping force have combined to shift the soliton to the very left edge of the pulse. In d(ii), the soliton is located at its intermediate trapping point, which is symmetrical with the pulse. In d(iii), the mismatch $d$ adds to the phase-based trapping force, causing the soliton to move closer to the peak. 
The observed change in noise transfer bandwidth with varied $d$ can be understood intuitively as being due to the local trapping gradient\cite{hendry_impact_2019-1} that gradually decreases from the edge of the pulse to the center. Analogous to atoms/particles trapped by optical potential wells\cite{phillips_nobel_1998}, a DKS trapped at a location closer to the input pulse center is subject to a shallower potential gradient, thus becoming more `free-running' and less affected by the noise contained in the driving field. For Fig. \ref{fig:simulation2}e(i) where the soliton is being `pulled' on the edge of the pulse, the broadening is maximized. For e(iv), where the soliton is instead being `pushed' near the peak of the pulse, it has almost reduced to zero. The reduced noise transfer effect is well reproduced by our simulation, as shown in Fig. \ref{fig:simulation2}(g). As $d$ increases, the DKS gets closer to the pulse center. Consequently the cut-off frequency of the noise transfer function decreases, showing remarkable agreement with the experimental measurement in Fig. \ref{fig:simulation2}(c).


\subsection*{\label{sec_summary}Summary}

Using the nonlinear, dispersion-engineered \SiN microresonator platform, we have generated a smooth, resonant supercontinuum based on dissipative soliton formation, comprising over 2,000 comb teeth. By exploiting the resonant enhancement of the high-$Q$ cavity, such a spectrum was generated with pulses 1-6 pJ in energy, $>$1 ps in duration, and on the order of single-Watt peak power. For future integration, the current EO-comb input could be replaced by an alternative provider of GHz rate, picosecond pulses, such as chip-based silicon or other semiconductor-based mode-locked lasers \cite{davenport_integrated_2018,delfyett_exploring_2019}. 
Further tailoring of the dispersion landscape and replacing the straight-waveguide coupling section with an adiabatic or curved coupling section \cite{chen_broadband_2017, moille_broadband_2019} will improve the generation and extraction of the short wavelength side of the spectrum. This way, the soliton comb bandwidth can be increased from 2/3rds of an octave to a full octave, enabling $f-2f$ self-referencing.
This work further invites a full exploration of the parameter space for input pulse chirp and flatness parameters in order to find further optimization of the frequency noise transfer, allowing the use of higher-noise voltage-controlled oscillators for locking the input pulse repetition rate.
Overall, this work demonstrates a new chip-based technique for direct access to broadband spectra at microwave repetition rates using a pulsed input, without the use of interleaving, or additional electro-optic modulation after the fact. Importantly, it can provide a way of balancing the fundamental efficiency restrictions between conventional supercontinuum generation, and dissipative soliton microcombs.

\subsection*{Acknowledgments}

This work was supported by Contract No. D18AC00032 (DRINQS) from the Defense Advanced Research Projects Agency (DARPA), and funding from the Swiss National Science Foundation under grant agreements No. 165933 and No. 176563. This material is based upon work supported by the Air Force Office of Scientific Research, Air Force Material Command, USAF under Award No. FA9550-15-1-0250. W. W. acknowledges support from the EU’s H2020 research and innovation programme under Marie Sklodowska-Curie IF grant agreement No. 753749 (SOLISYNTH).

\subsection*{Author Contributions}

M.H.A. performed the experiment with the assistance of R.B., E.O., and T.H.. The sample microresonator was fabricated by J.L.. M.H.A. conducted the numerical analysis. The manuscript was prepared by M.H.A. with the assistance of R.B., W.W., T.J.K., and T.H.. T.J.K. supervised the project.

\subsection*{Data Availability Statement} 

All data and analysis files will be made available via \texttt{zenodo.org} upon publication.

\subsection*{Methods} 

\begin{small}
	
\textbf{Electro-optic comb} The EO-comb is formed from a CW laser using an intensity modulator and three cascaded phase modulators, driven by an RF signal generator, creating approximately 50 spectral lines spaced by $f_\mathrm{eo}=$ 13.94 GHz. Two RF sources are used: RF-1: Rhode \& Schwarz SMB100A; RF-2:, Keysight E8267D. The dispersion-based waveform compression stage amounts to 300 m of standard SMF-28, plus an additional length from 5 m of dispersion-compensating fiber (DCF) in order to purposefully leave a residual positive chirp on the pulse, increasing pulse duration up to $\sim$1.4 ps when not amplified (see S.I.). 

\textbf{Microresonator} The \SiN microresonator used in this experiment has been fabricated with the \emph{photonic Damascene process} \cite{pfeiffer_photonic_2018} with a 2350$\times$770 nm$^2$ cross-section, and possesses a loaded $Q$ probability-distribution in the telecom band of $1.8\pm0.3\times10^6$. The cavity mode spectrum $\omega_\mu$, per mode index $\mu$, is expressed as $\omega_\mu-\omega_0-\mu D_1=\sum_{k\ge2}\mu^k\frac{D_k}{k!},\quad k\in \mathbb{N}$, with the higher-order cavity dispersion on the right-hand side. Spectroscopic measurements\cite{liu_frequency-comb-assisted_2016} yield $D_1=2\pi\cdot27.88$ GHz (the FSR), and $D_2=2\pi\cdot7.2$ kHz ($\beta_2=-2\pi D_2/LD_1^3$).
The location of the dispersive wave at $\omega_\mathrm{DW}=2\pi\cdot$154 THz allows us to infer $D_3\approx-3D_2D_1/(\omega_\mathrm{DW}-\omega_0)=2\pi\cdot15$ Hz ($\beta_3=-120$ fs$^3/$mm), when assuming $D_4=0$.

\textbf{DKS Measurement} 
For the calculations of the \emph{effective} power driving the resonant supercontinuum and conversion efficiency, we take into account the insertion loss of the lensed-fiber to chip interface (2.4 dB), and consider only every second EO-comb line, spaced by 28 GHz, as coupled into the microresonator. This further reduces the average power and conversion efficiency by 3 dB, and the effective pulse energy and pulse peak power seen by the resonator by 6 dB. The repetition rate beatnote, of the soliton comb lines only, was found by filtering out the EO-comb spectrum using a combination of a chirped fiber-Bragg grating and a wavelength-division multiplexer.

\textbf{Simulation method} For the noise-transfer numerical investigation, we use the LLE as our mean-field model without higher order perturbations such as third-order and higher-order dispersion, stimulated Raman scattering, and spectral $\kappa(\omega)$ response:

\begin{equation}
\label{eq:LLE1} 
\frac{\partial A(t)}{\partial t} = \left(  i\delta\omega -\frac{\kappa}{2} -i\frac{D_2}{2}\frac{\partial^2}{\partial\phi^2} + ig|A|^2 \right) A + \sqrt{\kappa_\mathrm{ex}}F(\phi,t)
\end{equation}

These omissions are so that the transfer of RF noise across the comb can be analyzed in its purest case. A full simulation of the generated DKS with all perturbations considered is provided in the S.I. The input pulse function $F(\phi,t)$ is expressed as a rectangular summation of lasing lines, similar to the experimental EO-comb:

\begin{equation}
	\label{eq:pulse}
	F(\phi,t) = \frac{\sqrt{P_0/\hbar\omega_0}}{M+1}\sum_{\mu=-M/2}^{M/2}e^{i\mu(\phi+\delta\phi(t))} e^{i\hat{D}_c}
\end{equation}

with $P_0$ the pulse peak power and $M+1$ the total number of laser lines coupled to the cavity with spacing equal to the FSR. A residual dispersion is also applied to the pulse spectrum $\hat{D}_c$, which expresses the inexact dispersion compensation compressing the experimental EO-comb waveform (see S.I.). The long-term phase noise noise of the driving pulse is provided by $\delta\phi(t)$, related to the frequency noise of the RF source by $\delta\phi(t)=\int^{t}2\pi \,\delta\!f(t')dt'$.

The simulation parameters are chosen to reflect a basic \SiN resonator similar to that of the experiment: $\kappa=2\pi\cdot100$ MHz, coupling rate $\kappa_\mathrm{ex}=2\pi\cdot50$ MHz (critical coupling), dispersion $D_2=2\pi\cdot28$ kHz, and nonlinear coupling $g=2\pi\cdot0.054$ Hz. The driving parameters are $P_0=900$ mW (24$\times$ parametric oscillation threshold), $\delta\omega=6\kappa$, and $M=24$. Additional normal dispersion $\beta_c=+0.3$ ps$^2$, giving a spectral phase profile $\hat{D}_c=\beta_c(\mu D_1)^2/2$. Frequency-noise power spectra are found by Fourier-transform of the slow-time phase fluctuations of individual comb lines $\tilde{A}_\mu(t)$, as $\delta\!f_\mu(t) = \frac{d}{dt} \arg(\tilde{A}_\mu(t))$. 

As according to \cite{hendry_spontaneous_2018}, the critical amplitude $F_C$ to which a soliton locks for this detuning ($\delta\omega>3\kappa$), based on pure intensity-based trapping, is close to the minimum amplitude required for DKS existence $|F_c|^2 \geq \frac{2\kappa^2\delta\omega}{\pi^2g\kappa_\mathrm{ex}}$.

\end{small}

\bibliographystyle{apsrev4-2}
\bibliography{biblib_2}

\begin{thebibliography}{73}%
\makeatletter
\providecommand \@ifxundefined [1]{%
 \@ifx{#1\undefined}
}%
\providecommand \@ifnum [1]{%
 \ifnum #1\expandafter \@firstoftwo
 \else \expandafter \@secondoftwo
 \fi
}%
\providecommand \@ifx [1]{%
 \ifx #1\expandafter \@firstoftwo
 \else \expandafter \@secondoftwo
 \fi
}%
\providecommand \natexlab [1]{#1}%
\providecommand \enquote  [1]{``#1''}%
\providecommand \bibnamefont  [1]{#1}%
\providecommand \bibfnamefont [1]{#1}%
\providecommand \citenamefont [1]{#1}%
\providecommand \href@noop [0]{\@secondoftwo}%
\providecommand \href [0]{\begingroup \@sanitize@url \@href}%
\providecommand \@href[1]{\@@startlink{#1}\@@href}%
\providecommand \@@href[1]{\endgroup#1\@@endlink}%
\providecommand \@sanitize@url [0]{\catcode `\\12\catcode `\$12\catcode
  `\&12\catcode `\#12\catcode `\^12\catcode `\_12\catcode `\%12\relax}%
\providecommand \@@startlink[1]{}%
\providecommand \@@endlink[0]{}%
\providecommand \url  [0]{\begingroup\@sanitize@url \@url }%
\providecommand \@url [1]{\endgroup\@href {#1}{\urlprefix }}%
\providecommand \urlprefix  [0]{URL }%
\providecommand \Eprint [0]{\href }%
\providecommand \doibase [0]{https://doi.org/}%
\providecommand \selectlanguage [0]{\@gobble}%
\providecommand \bibinfo  [0]{\@secondoftwo}%
\providecommand \bibfield  [0]{\@secondoftwo}%
\providecommand \translation [1]{[#1]}%
\providecommand \BibitemOpen [0]{}%
\providecommand \bibitemStop [0]{}%
\providecommand \bibitemNoStop [0]{.\EOS\space}%
\providecommand \EOS [0]{\spacefactor3000\relax}%
\providecommand \BibitemShut  [1]{\csname bibitem#1\endcsname}%
\let\auto@bib@innerbib\@empty
\bibitem [{\citenamefont {Alfano}\ and\ \citenamefont
  {Shapiro}(1970)}]{alfano_observation_1970}%
  \BibitemOpen
  \bibfield  {author} {\bibinfo {author} {\bibfnamefont {R.~R.}\ \bibnamefont
  {Alfano}}\ and\ \bibinfo {author} {\bibfnamefont {S.~L.}\ \bibnamefont
  {Shapiro}},\ }\href {https://doi.org/10.1103/PhysRevLett.24.592} {\bibfield
  {journal} {\bibinfo  {journal} {Physical Review Letters}\ }\textbf {\bibinfo
  {volume} {24}},\ \bibinfo {pages} {592} (\bibinfo {year} {1970})}\BibitemShut
  {NoStop}%
\bibitem [{\citenamefont {Ranka}\ \emph {et~al.}(2000)\citenamefont {Ranka},
  \citenamefont {Windeler},\ and\ \citenamefont {Stentz}}]{ranka_visible_2000}%
  \BibitemOpen
  \bibfield  {author} {\bibinfo {author} {\bibfnamefont {J.~K.}\ \bibnamefont
  {Ranka}}, \bibinfo {author} {\bibfnamefont {R.~S.}\ \bibnamefont
  {Windeler}},\ and\ \bibinfo {author} {\bibfnamefont {A.~J.}\ \bibnamefont
  {Stentz}},\ }\href {https://doi.org/10.1364/OL.25.000025} {\bibfield
  {journal} {\bibinfo  {journal} {Optics Letters}\ }\textbf {\bibinfo {volume}
  {25}},\ \bibinfo {pages} {25} (\bibinfo {year} {2000})}\BibitemShut {NoStop}%
\bibitem [{\citenamefont {Birks}\ \emph {et~al.}(2000)\citenamefont {Birks},
  \citenamefont {Wadsworth},\ and\ \citenamefont
  {Russell}}]{birks_supercontinuum_2000}%
  \BibitemOpen
  \bibfield  {author} {\bibinfo {author} {\bibfnamefont {T.~A.}\ \bibnamefont
  {Birks}}, \bibinfo {author} {\bibfnamefont {W.~J.}\ \bibnamefont
  {Wadsworth}},\ and\ \bibinfo {author} {\bibfnamefont {P.~S.~J.}\ \bibnamefont
  {Russell}},\ }\href {https://doi.org/10.1364/OL.25.001415} {\bibfield
  {journal} {\bibinfo  {journal} {Optics Letters}\ }\textbf {\bibinfo {volume}
  {25}},\ \bibinfo {pages} {1415} (\bibinfo {year} {2000})}\BibitemShut
  {NoStop}%
\bibitem [{\citenamefont {Jones}\ \emph {et~al.}(2000)\citenamefont {Jones},
  \citenamefont {Diddams}, \citenamefont {Ranka}, \citenamefont {Stentz},
  \citenamefont {Windeler}, \citenamefont {Hall},\ and\ \citenamefont
  {Cundiff}}]{jones_carrier-envelope_2000}%
  \BibitemOpen
  \bibfield  {author} {\bibinfo {author} {\bibfnamefont {D.~J.}\ \bibnamefont
  {Jones}}, \bibinfo {author} {\bibfnamefont {S.~A.}\ \bibnamefont {Diddams}},
  \bibinfo {author} {\bibfnamefont {J.~K.}\ \bibnamefont {Ranka}}, \bibinfo
  {author} {\bibfnamefont {A.}~\bibnamefont {Stentz}}, \bibinfo {author}
  {\bibfnamefont {R.~S.}\ \bibnamefont {Windeler}}, \bibinfo {author}
  {\bibfnamefont {J.~L.}\ \bibnamefont {Hall}},\ and\ \bibinfo {author}
  {\bibfnamefont {S.~T.}\ \bibnamefont {Cundiff}},\ }\href
  {https://doi.org/10.1126/science.288.5466.635} {\bibfield  {journal}
  {\bibinfo  {journal} {Science}\ }\textbf {\bibinfo {volume} {288}},\ \bibinfo
  {pages} {635} (\bibinfo {year} {2000})}\BibitemShut {NoStop}%
\bibitem [{\citenamefont {Udem}\ \emph {et~al.}(2002)\citenamefont {Udem},
  \citenamefont {Holzwarth},\ and\ \citenamefont
  {H\"ansch}}]{udem_optical_2002}%
  \BibitemOpen
  \bibfield  {author} {\bibinfo {author} {\bibfnamefont {T.}~\bibnamefont
  {Udem}}, \bibinfo {author} {\bibfnamefont {R.}~\bibnamefont {Holzwarth}},\
  and\ \bibinfo {author} {\bibfnamefont {T.~W.}\ \bibnamefont {H\"ansch}},\
  }\href {https://doi.org/10.1038/416233a} {\bibfield  {journal} {\bibinfo
  {journal} {Nature}\ }\textbf {\bibinfo {volume} {416}},\ \bibinfo {pages}
  {233} (\bibinfo {year} {2002})}\BibitemShut {NoStop}%
\bibitem [{\citenamefont {Nakazawa}\ \emph {et~al.}(1998)\citenamefont
  {Nakazawa}, \citenamefont {Tamura}, \citenamefont {Kubota},\ and\
  \citenamefont {Yoshida}}]{nakazawa_coherence_1998}%
  \BibitemOpen
  \bibfield  {author} {\bibinfo {author} {\bibfnamefont {M.}~\bibnamefont
  {Nakazawa}}, \bibinfo {author} {\bibfnamefont {K.}~\bibnamefont {Tamura}},
  \bibinfo {author} {\bibfnamefont {H.}~\bibnamefont {Kubota}},\ and\ \bibinfo
  {author} {\bibfnamefont {E.}~\bibnamefont {Yoshida}},\ }\href
  {https://doi.org/10.1006/ofte.1998.0253} {\bibfield  {journal} {\bibinfo
  {journal} {Optical Fiber Technology}\ }\textbf {\bibinfo {volume} {4}},\
  \bibinfo {pages} {215} (\bibinfo {year} {1998})}\BibitemShut {NoStop}%
\bibitem [{\citenamefont {Dudley}\ \emph {et~al.}(2006)\citenamefont {Dudley},
  \citenamefont {Genty},\ and\ \citenamefont
  {Coen}}]{dudley_supercontinuum_2006}%
  \BibitemOpen
  \bibfield  {author} {\bibinfo {author} {\bibfnamefont {J.~M.}\ \bibnamefont
  {Dudley}}, \bibinfo {author} {\bibfnamefont {G.}~\bibnamefont {Genty}},\ and\
  \bibinfo {author} {\bibfnamefont {S.}~\bibnamefont {Coen}},\ }\href
  {https://doi.org/10.1103/RevModPhys.78.1135} {\bibfield  {journal} {\bibinfo
  {journal} {Reviews of Modern Physics}\ }\textbf {\bibinfo {volume} {78}},\
  \bibinfo {pages} {1135} (\bibinfo {year} {2006})}\BibitemShut {NoStop}%
\bibitem [{\citenamefont {Xie}\ \emph {et~al.}(2017)\citenamefont {Xie},
  \citenamefont {Bouchand}, \citenamefont {Nicolodi}, \citenamefont {Giunta},
  \citenamefont {H\"ansel}, \citenamefont {Lezius}, \citenamefont {Joshi},
  \citenamefont {Datta}, \citenamefont {Alexandre}, \citenamefont {Lours},
  \citenamefont {Tremblin}, \citenamefont {Santarelli}, \citenamefont
  {Holzwarth},\ and\ \citenamefont {Le~Coq}}]{xie_photonic_2017}%
  \BibitemOpen
  \bibfield  {author} {\bibinfo {author} {\bibfnamefont {X.}~\bibnamefont
  {Xie}}, \bibinfo {author} {\bibfnamefont {R.}~\bibnamefont {Bouchand}},
  \bibinfo {author} {\bibfnamefont {D.}~\bibnamefont {Nicolodi}}, \bibinfo
  {author} {\bibfnamefont {M.}~\bibnamefont {Giunta}}, \bibinfo {author}
  {\bibfnamefont {W.}~\bibnamefont {H\"ansel}}, \bibinfo {author}
  {\bibfnamefont {M.}~\bibnamefont {Lezius}}, \bibinfo {author} {\bibfnamefont
  {A.}~\bibnamefont {Joshi}}, \bibinfo {author} {\bibfnamefont
  {S.}~\bibnamefont {Datta}}, \bibinfo {author} {\bibfnamefont
  {C.}~\bibnamefont {Alexandre}}, \bibinfo {author} {\bibfnamefont
  {M.}~\bibnamefont {Lours}}, \bibinfo {author} {\bibfnamefont {P.-A.}\
  \bibnamefont {Tremblin}}, \bibinfo {author} {\bibfnamefont {G.}~\bibnamefont
  {Santarelli}}, \bibinfo {author} {\bibfnamefont {R.}~\bibnamefont
  {Holzwarth}},\ and\ \bibinfo {author} {\bibfnamefont {Y.}~\bibnamefont
  {Le~Coq}},\ }\href {https://doi.org/10.1038/nphoton.2016.215} {\bibfield
  {journal} {\bibinfo  {journal} {Nature Photonics}\ }\textbf {\bibinfo
  {volume} {11}},\ \bibinfo {pages} {44} (\bibinfo {year} {2017})}\BibitemShut
  {NoStop}%
\bibitem [{\citenamefont {Ideguchi}\ \emph {et~al.}(2013)\citenamefont
  {Ideguchi}, \citenamefont {Holzner}, \citenamefont {Bernhardt}, \citenamefont
  {Guelachvili}, \citenamefont {Picqu\'e},\ and\ \citenamefont
  {H\"ansch}}]{ideguchi_coherent_2013}%
  \BibitemOpen
  \bibfield  {author} {\bibinfo {author} {\bibfnamefont {T.}~\bibnamefont
  {Ideguchi}}, \bibinfo {author} {\bibfnamefont {S.}~\bibnamefont {Holzner}},
  \bibinfo {author} {\bibfnamefont {B.}~\bibnamefont {Bernhardt}}, \bibinfo
  {author} {\bibfnamefont {G.}~\bibnamefont {Guelachvili}}, \bibinfo {author}
  {\bibfnamefont {N.}~\bibnamefont {Picqu\'e}},\ and\ \bibinfo {author}
  {\bibfnamefont {T.~W.}\ \bibnamefont {H\"ansch}},\ }\href
  {https://doi.org/10.1038/nature12607} {\bibfield  {journal} {\bibinfo
  {journal} {Nature}\ }\textbf {\bibinfo {volume} {502}},\ \bibinfo {pages}
  {355} (\bibinfo {year} {2013})}\BibitemShut {NoStop}%
\bibitem [{\citenamefont {Marin-Palomo}\ \emph {et~al.}(2017)\citenamefont
  {Marin-Palomo}, \citenamefont {Kemal}, \citenamefont {Karpov}, \citenamefont
  {Kordts}, \citenamefont {Pfeifle}, \citenamefont {Pfeiffer}, \citenamefont
  {Trocha}, \citenamefont {Wolf}, \citenamefont {Brasch}, \citenamefont
  {Anderson}, \citenamefont {Rosenberger}, \citenamefont {Vijayan},
  \citenamefont {Freude}, \citenamefont {Kippenberg},\ and\ \citenamefont
  {Koos}}]{marin-palomo_microresonator-based_2017}%
  \BibitemOpen
  \bibfield  {author} {\bibinfo {author} {\bibfnamefont {P.}~\bibnamefont
  {Marin-Palomo}}, \bibinfo {author} {\bibfnamefont {J.~N.}\ \bibnamefont
  {Kemal}}, \bibinfo {author} {\bibfnamefont {M.}~\bibnamefont {Karpov}},
  \bibinfo {author} {\bibfnamefont {A.}~\bibnamefont {Kordts}}, \bibinfo
  {author} {\bibfnamefont {J.}~\bibnamefont {Pfeifle}}, \bibinfo {author}
  {\bibfnamefont {M.~H.~P.}\ \bibnamefont {Pfeiffer}}, \bibinfo {author}
  {\bibfnamefont {P.}~\bibnamefont {Trocha}}, \bibinfo {author} {\bibfnamefont
  {S.}~\bibnamefont {Wolf}}, \bibinfo {author} {\bibfnamefont {V.}~\bibnamefont
  {Brasch}}, \bibinfo {author} {\bibfnamefont {M.~H.}\ \bibnamefont
  {Anderson}}, \bibinfo {author} {\bibfnamefont {R.}~\bibnamefont
  {Rosenberger}}, \bibinfo {author} {\bibfnamefont {K.}~\bibnamefont
  {Vijayan}}, \bibinfo {author} {\bibfnamefont {W.}~\bibnamefont {Freude}},
  \bibinfo {author} {\bibfnamefont {T.~J.}\ \bibnamefont {Kippenberg}},\ and\
  \bibinfo {author} {\bibfnamefont {C.}~\bibnamefont {Koos}},\ }\href
  {https://doi.org/10.1038/nature22387} {\bibfield  {journal} {\bibinfo
  {journal} {Nature}\ }\textbf {\bibinfo {volume} {546}},\ \bibinfo {pages}
  {274} (\bibinfo {year} {2017})}\BibitemShut {NoStop}%
\bibitem [{\citenamefont {Murphy}\ \emph {et~al.}(2007)\citenamefont {Murphy},
  \citenamefont {Udem}, \citenamefont {Holzwarth}, \citenamefont {Sizmann},
  \citenamefont {Pasquini}, \citenamefont {Araujo-Hauck}, \citenamefont
  {Dekker}, \citenamefont {D'Odorico}, \citenamefont {Fischer}, \citenamefont
  {H\"ansch},\ and\ \citenamefont {Manescau}}]{murphy_high-precision_2007}%
  \BibitemOpen
  \bibfield  {author} {\bibinfo {author} {\bibfnamefont {M.~T.}\ \bibnamefont
  {Murphy}}, \bibinfo {author} {\bibfnamefont {T.}~\bibnamefont {Udem}},
  \bibinfo {author} {\bibfnamefont {R.}~\bibnamefont {Holzwarth}}, \bibinfo
  {author} {\bibfnamefont {A.}~\bibnamefont {Sizmann}}, \bibinfo {author}
  {\bibfnamefont {L.}~\bibnamefont {Pasquini}}, \bibinfo {author}
  {\bibfnamefont {C.}~\bibnamefont {Araujo-Hauck}}, \bibinfo {author}
  {\bibfnamefont {H.}~\bibnamefont {Dekker}}, \bibinfo {author} {\bibfnamefont
  {S.}~\bibnamefont {D'Odorico}}, \bibinfo {author} {\bibfnamefont
  {M.}~\bibnamefont {Fischer}}, \bibinfo {author} {\bibfnamefont {T.~W.}\
  \bibnamefont {H\"ansch}},\ and\ \bibinfo {author} {\bibfnamefont
  {A.}~\bibnamefont {Manescau}},\ }\href
  {https://doi.org/10.1111/j.1365-2966.2007.12147.x} {\bibfield  {journal}
  {\bibinfo  {journal} {Monthly Notices of the Royal Astronomical Society}\
  }\textbf {\bibinfo {volume} {380}},\ \bibinfo {pages} {839} (\bibinfo {year}
  {2007})}\BibitemShut {NoStop}%
\bibitem [{\citenamefont {Herr}\ \emph {et~al.}(2014)\citenamefont {Herr},
  \citenamefont {Brasch}, \citenamefont {Jost}, \citenamefont {Wang},
  \citenamefont {Kondratiev}, \citenamefont {Gorodetsky},\ and\ \citenamefont
  {Kippenberg}}]{herr_temporal_2014}%
  \BibitemOpen
  \bibfield  {author} {\bibinfo {author} {\bibfnamefont {T.}~\bibnamefont
  {Herr}}, \bibinfo {author} {\bibfnamefont {V.}~\bibnamefont {Brasch}},
  \bibinfo {author} {\bibfnamefont {J.~D.}\ \bibnamefont {Jost}}, \bibinfo
  {author} {\bibfnamefont {C.~Y.}\ \bibnamefont {Wang}}, \bibinfo {author}
  {\bibfnamefont {N.~M.}\ \bibnamefont {Kondratiev}}, \bibinfo {author}
  {\bibfnamefont {M.~L.}\ \bibnamefont {Gorodetsky}},\ and\ \bibinfo {author}
  {\bibfnamefont {T.~J.}\ \bibnamefont {Kippenberg}},\ }\href
  {https://doi.org/10.1038/nphoton.2013.343} {\bibfield  {journal} {\bibinfo
  {journal} {Nature Photonics}\ }\textbf {\bibinfo {volume} {8}},\ \bibinfo
  {pages} {145} (\bibinfo {year} {2014})}\BibitemShut {NoStop}%
\bibitem [{\citenamefont {Kippenberg}\ \emph {et~al.}(2018)\citenamefont
  {Kippenberg}, \citenamefont {Gaeta}, \citenamefont {Lipson},\ and\
  \citenamefont {Gorodetsky}}]{kippenberg_dissipative_2018}%
  \BibitemOpen
  \bibfield  {author} {\bibinfo {author} {\bibfnamefont {T.~J.}\ \bibnamefont
  {Kippenberg}}, \bibinfo {author} {\bibfnamefont {A.~L.}\ \bibnamefont
  {Gaeta}}, \bibinfo {author} {\bibfnamefont {M.}~\bibnamefont {Lipson}},\ and\
  \bibinfo {author} {\bibfnamefont {M.~L.}\ \bibnamefont {Gorodetsky}},\ }\href
  {https://doi.org/10.1126/science.aan8083} {\bibfield  {journal} {\bibinfo
  {journal} {Science}\ }\textbf {\bibinfo {volume} {361}},\ \bibinfo {pages}
  {eaan8083} (\bibinfo {year} {2018})}\BibitemShut {NoStop}%
\bibitem [{\citenamefont {Li}\ \emph {et~al.}(2017)\citenamefont {Li},
  \citenamefont {Briles}, \citenamefont {Westly}, \citenamefont {Drake},
  \citenamefont {Stone}, \citenamefont {Ilic}, \citenamefont {Diddams},
  \citenamefont {Papp},\ and\ \citenamefont {Srinivasan}}]{li_stably_2017}%
  \BibitemOpen
  \bibfield  {author} {\bibinfo {author} {\bibfnamefont {Q.}~\bibnamefont
  {Li}}, \bibinfo {author} {\bibfnamefont {T.~C.}\ \bibnamefont {Briles}},
  \bibinfo {author} {\bibfnamefont {D.~A.}\ \bibnamefont {Westly}}, \bibinfo
  {author} {\bibfnamefont {T.~E.}\ \bibnamefont {Drake}}, \bibinfo {author}
  {\bibfnamefont {J.~R.}\ \bibnamefont {Stone}}, \bibinfo {author}
  {\bibfnamefont {B.~R.}\ \bibnamefont {Ilic}}, \bibinfo {author}
  {\bibfnamefont {S.~A.}\ \bibnamefont {Diddams}}, \bibinfo {author}
  {\bibfnamefont {S.~B.}\ \bibnamefont {Papp}},\ and\ \bibinfo {author}
  {\bibfnamefont {K.}~\bibnamefont {Srinivasan}},\ }\href
  {https://doi.org/10.1364/OPTICA.4.000193} {\bibfield  {journal} {\bibinfo
  {journal} {Optica}\ }\textbf {\bibinfo {volume} {4}},\ \bibinfo {pages} {193}
  (\bibinfo {year} {2017})}\BibitemShut {NoStop}%
\bibitem [{\citenamefont {Pfeiffer}\ \emph {et~al.}(2017)\citenamefont
  {Pfeiffer}, \citenamefont {Herkommer}, \citenamefont {Liu}, \citenamefont
  {Guo}, \citenamefont {Karpov}, \citenamefont {Lucas}, \citenamefont
  {Zervas},\ and\ \citenamefont {Kippenberg}}]{pfeiffer_octave-spanning_2017}%
  \BibitemOpen
  \bibfield  {author} {\bibinfo {author} {\bibfnamefont {M.~H.~P.}\
  \bibnamefont {Pfeiffer}}, \bibinfo {author} {\bibfnamefont {C.}~\bibnamefont
  {Herkommer}}, \bibinfo {author} {\bibfnamefont {J.}~\bibnamefont {Liu}},
  \bibinfo {author} {\bibfnamefont {H.}~\bibnamefont {Guo}}, \bibinfo {author}
  {\bibfnamefont {M.}~\bibnamefont {Karpov}}, \bibinfo {author} {\bibfnamefont
  {E.}~\bibnamefont {Lucas}}, \bibinfo {author} {\bibfnamefont
  {M.}~\bibnamefont {Zervas}},\ and\ \bibinfo {author} {\bibfnamefont {T.~J.}\
  \bibnamefont {Kippenberg}},\ }\href {https://doi.org/10.1364/OPTICA.4.000684}
  {\bibfield  {journal} {\bibinfo  {journal} {Optica}\ }\textbf {\bibinfo
  {volume} {4}},\ \bibinfo {pages} {684} (\bibinfo {year} {2017})}\BibitemShut
  {NoStop}%
\bibitem [{\citenamefont {Bao}\ \emph {et~al.}(2014)\citenamefont {Bao},
  \citenamefont {Zhang}, \citenamefont {Matsko}, \citenamefont {Yan},
  \citenamefont {Zhao}, \citenamefont {Xie}, \citenamefont {Agarwal},
  \citenamefont {Kimerling}, \citenamefont {Michel}, \citenamefont {Maleki},\
  and\ \citenamefont {Willner}}]{bao_nonlinear_2014}%
  \BibitemOpen
  \bibfield  {author} {\bibinfo {author} {\bibfnamefont {C.}~\bibnamefont
  {Bao}}, \bibinfo {author} {\bibfnamefont {L.}~\bibnamefont {Zhang}}, \bibinfo
  {author} {\bibfnamefont {A.}~\bibnamefont {Matsko}}, \bibinfo {author}
  {\bibfnamefont {Y.}~\bibnamefont {Yan}}, \bibinfo {author} {\bibfnamefont
  {Z.}~\bibnamefont {Zhao}}, \bibinfo {author} {\bibfnamefont {G.}~\bibnamefont
  {Xie}}, \bibinfo {author} {\bibfnamefont {A.~M.}\ \bibnamefont {Agarwal}},
  \bibinfo {author} {\bibfnamefont {L.~C.}\ \bibnamefont {Kimerling}}, \bibinfo
  {author} {\bibfnamefont {J.}~\bibnamefont {Michel}}, \bibinfo {author}
  {\bibfnamefont {L.}~\bibnamefont {Maleki}},\ and\ \bibinfo {author}
  {\bibfnamefont {A.~E.}\ \bibnamefont {Willner}},\ }\href
  {https://doi.org/10.1364/OL.39.006126} {\bibfield  {journal} {\bibinfo
  {journal} {Optics Letters}\ }\textbf {\bibinfo {volume} {39}},\ \bibinfo
  {pages} {6126} (\bibinfo {year} {2014})}\BibitemShut {NoStop}%
\bibitem [{\citenamefont {Gaeta}\ \emph {et~al.}(2019)\citenamefont {Gaeta},
  \citenamefont {Lipson},\ and\ \citenamefont
  {Kippenberg}}]{gaeta_photonic-chip-based_2019}%
  \BibitemOpen
  \bibfield  {author} {\bibinfo {author} {\bibfnamefont {A.~L.}\ \bibnamefont
  {Gaeta}}, \bibinfo {author} {\bibfnamefont {M.}~\bibnamefont {Lipson}},\ and\
  \bibinfo {author} {\bibfnamefont {T.~J.}\ \bibnamefont {Kippenberg}},\ }\href
  {https://doi.org/10.1038/s41566-019-0358-x} {\bibfield  {journal} {\bibinfo
  {journal} {Nature Photonics}\ }\textbf {\bibinfo {volume} {13}},\ \bibinfo
  {pages} {158} (\bibinfo {year} {2019})}\BibitemShut {NoStop}%
\bibitem [{\citenamefont {Obrzud}\ \emph {et~al.}(2017)\citenamefont {Obrzud},
  \citenamefont {Lecomte},\ and\ \citenamefont {Herr}}]{obrzud_temporal_2017}%
  \BibitemOpen
  \bibfield  {author} {\bibinfo {author} {\bibfnamefont {E.}~\bibnamefont
  {Obrzud}}, \bibinfo {author} {\bibfnamefont {S.}~\bibnamefont {Lecomte}},\
  and\ \bibinfo {author} {\bibfnamefont {T.}~\bibnamefont {Herr}},\ }\href
  {https://doi.org/10.1038/nphoton.2017.140} {\bibfield  {journal} {\bibinfo
  {journal} {Nature Photonics}\ }\textbf {\bibinfo {volume} {11}},\ \bibinfo
  {pages} {nphoton.2017.140} (\bibinfo {year} {2017})}\BibitemShut {NoStop}%
\bibitem [{\citenamefont {Liu}\ \emph {et~al.}(2018)\citenamefont {Liu},
  \citenamefont {Raja}, \citenamefont {Karpov}, \citenamefont {Ghadiani},
  \citenamefont {Pfeiffer}, \citenamefont {Du}, \citenamefont {Engelsen},
  \citenamefont {Guo}, \citenamefont {Zervas},\ and\ \citenamefont
  {Kippenberg}}]{liu_ultralow-power_2018-1}%
  \BibitemOpen
  \bibfield  {author} {\bibinfo {author} {\bibfnamefont {J.}~\bibnamefont
  {Liu}}, \bibinfo {author} {\bibfnamefont {A.~S.}\ \bibnamefont {Raja}},
  \bibinfo {author} {\bibfnamefont {M.}~\bibnamefont {Karpov}}, \bibinfo
  {author} {\bibfnamefont {B.}~\bibnamefont {Ghadiani}}, \bibinfo {author}
  {\bibfnamefont {M.~H.~P.}\ \bibnamefont {Pfeiffer}}, \bibinfo {author}
  {\bibfnamefont {B.}~\bibnamefont {Du}}, \bibinfo {author} {\bibfnamefont
  {N.~J.}\ \bibnamefont {Engelsen}}, \bibinfo {author} {\bibfnamefont
  {H.}~\bibnamefont {Guo}}, \bibinfo {author} {\bibfnamefont {M.}~\bibnamefont
  {Zervas}},\ and\ \bibinfo {author} {\bibfnamefont {T.~J.}\ \bibnamefont
  {Kippenberg}},\ }\href {https://doi.org/10.1364/OPTICA.5.001347} {\bibfield
  {journal} {\bibinfo  {journal} {Optica}\ }\textbf {\bibinfo {volume} {5}},\
  \bibinfo {pages} {1347} (\bibinfo {year} {2018})}\BibitemShut {NoStop}%
\bibitem [{\citenamefont {Weng}\ \emph {et~al.}(2019)\citenamefont {Weng},
  \citenamefont {Lucas}, \citenamefont {Lihachev}, \citenamefont {Lobanov},
  \citenamefont {Guo}, \citenamefont {Gorodetsky},\ and\ \citenamefont
  {Kippenberg}}]{weng_spectral_2019}%
  \BibitemOpen
  \bibfield  {author} {\bibinfo {author} {\bibfnamefont {W.}~\bibnamefont
  {Weng}}, \bibinfo {author} {\bibfnamefont {E.}~\bibnamefont {Lucas}},
  \bibinfo {author} {\bibfnamefont {G.}~\bibnamefont {Lihachev}}, \bibinfo
  {author} {\bibfnamefont {V.~E.}\ \bibnamefont {Lobanov}}, \bibinfo {author}
  {\bibfnamefont {H.}~\bibnamefont {Guo}}, \bibinfo {author} {\bibfnamefont
  {M.~L.}\ \bibnamefont {Gorodetsky}},\ and\ \bibinfo {author} {\bibfnamefont
  {T.~J.}\ \bibnamefont {Kippenberg}},\ }\href
  {https://doi.org/10.1103/PhysRevLett.122.013902} {\bibfield  {journal}
  {\bibinfo  {journal} {Physical Review Letters}\ }\textbf {\bibinfo {volume}
  {122}},\ \bibinfo {pages} {013902} (\bibinfo {year} {2019})}\BibitemShut
  {NoStop}%
\bibitem [{\citenamefont {Brasch}\ \emph {et~al.}(2019)\citenamefont {Brasch},
  \citenamefont {Obrzud}, \citenamefont {Lecomte},\ and\ \citenamefont
  {Herr}}]{brasch_nonlinear_2019}%
  \BibitemOpen
  \bibfield  {author} {\bibinfo {author} {\bibfnamefont {V.}~\bibnamefont
  {Brasch}}, \bibinfo {author} {\bibfnamefont {E.}~\bibnamefont {Obrzud}},
  \bibinfo {author} {\bibfnamefont {S.}~\bibnamefont {Lecomte}},\ and\ \bibinfo
  {author} {\bibfnamefont {T.}~\bibnamefont {Herr}},\ }\href
  {http://arxiv.org/abs/1907.09715} {\bibfield  {journal} {\bibinfo  {journal}
  {arXiv:1907.09715 [physics]}\ } (\bibinfo {year} {2019})},\ \bibinfo {note}
  {arXiv: 1907.09715}\BibitemShut {NoStop}%
\bibitem [{\citenamefont {Bellini}\ and\ \citenamefont
  {H\"ansch}(2000)}]{bellini_phase-locked_2000}%
  \BibitemOpen
  \bibfield  {author} {\bibinfo {author} {\bibfnamefont {M.}~\bibnamefont
  {Bellini}}\ and\ \bibinfo {author} {\bibfnamefont {T.~W.}\ \bibnamefont
  {H\"ansch}},\ }\href {https://doi.org/10.1364/OL.25.001049} {\bibfield
  {journal} {\bibinfo  {journal} {Optics Letters}\ }\textbf {\bibinfo {volume}
  {25}},\ \bibinfo {pages} {1049} (\bibinfo {year} {2000})}\BibitemShut
  {NoStop}%
\bibitem [{\citenamefont {Russell}(2006)}]{russell_photonic-crystal_2006}%
  \BibitemOpen
  \bibfield  {author} {\bibinfo {author} {\bibfnamefont {P.~S.~J.}\
  \bibnamefont {Russell}},\ }\href
  {https://www.osapublishing.org/jlt/abstract.cfm?uri=jlt-24-12-4729}
  {\bibfield  {journal} {\bibinfo  {journal} {Journal of Lightwave Technology}\
  }\textbf {\bibinfo {volume} {24}},\ \bibinfo {pages} {4729} (\bibinfo {year}
  {2006})}\BibitemShut {NoStop}%
\bibitem [{\citenamefont {Skryabin}\ and\ \citenamefont
  {Gorbach}(2010)}]{skryabin_colloquium:_2010}%
  \BibitemOpen
  \bibfield  {author} {\bibinfo {author} {\bibfnamefont {D.~V.}\ \bibnamefont
  {Skryabin}}\ and\ \bibinfo {author} {\bibfnamefont {A.~V.}\ \bibnamefont
  {Gorbach}},\ }\href {https://doi.org/10.1103/RevModPhys.82.1287} {\bibfield
  {journal} {\bibinfo  {journal} {Reviews of Modern Physics}\ }\textbf
  {\bibinfo {volume} {82}},\ \bibinfo {pages} {1287} (\bibinfo {year}
  {2010})}\BibitemShut {NoStop}%
\bibitem [{\citenamefont {Herrmann}\ \emph {et~al.}(2002)\citenamefont
  {Herrmann}, \citenamefont {Griebner}, \citenamefont {Zhavoronkov},
  \citenamefont {Husakou}, \citenamefont {Nickel}, \citenamefont {Knight},
  \citenamefont {Wadsworth}, \citenamefont {Russell},\ and\ \citenamefont
  {Korn}}]{herrmann_experimental_2002}%
  \BibitemOpen
  \bibfield  {author} {\bibinfo {author} {\bibfnamefont {J.}~\bibnamefont
  {Herrmann}}, \bibinfo {author} {\bibfnamefont {U.}~\bibnamefont {Griebner}},
  \bibinfo {author} {\bibfnamefont {N.}~\bibnamefont {Zhavoronkov}}, \bibinfo
  {author} {\bibfnamefont {A.}~\bibnamefont {Husakou}}, \bibinfo {author}
  {\bibfnamefont {D.}~\bibnamefont {Nickel}}, \bibinfo {author} {\bibfnamefont
  {J.~C.}\ \bibnamefont {Knight}}, \bibinfo {author} {\bibfnamefont {W.~J.}\
  \bibnamefont {Wadsworth}}, \bibinfo {author} {\bibfnamefont {P.~S.~J.}\
  \bibnamefont {Russell}},\ and\ \bibinfo {author} {\bibfnamefont
  {G.}~\bibnamefont {Korn}},\ }\bibfield  {journal} {\bibinfo  {journal}
  {Physical Review Letters}\ }\textbf {\bibinfo {volume} {88}},\ \href
  {https://doi.org/10.1103/PhysRevLett.88.173901}
  {10.1103/PhysRevLett.88.173901} (\bibinfo {year} {2002})\BibitemShut
  {NoStop}%
\bibitem [{\citenamefont {Akhmediev}\ and\ \citenamefont
  {Karlsson}(1995)}]{akhmediev_cherenkov_1995}%
  \BibitemOpen
  \bibfield  {author} {\bibinfo {author} {\bibfnamefont {N.}~\bibnamefont
  {Akhmediev}}\ and\ \bibinfo {author} {\bibfnamefont {M.}~\bibnamefont
  {Karlsson}},\ }\href {https://doi.org/10.1103/PhysRevA.51.2602} {\bibfield
  {journal} {\bibinfo  {journal} {Physical Review A}\ }\textbf {\bibinfo
  {volume} {51}},\ \bibinfo {pages} {2602} (\bibinfo {year}
  {1995})}\BibitemShut {NoStop}%
\bibitem [{\citenamefont {Hilligsøe}\ \emph {et~al.}(2003)\citenamefont
  {Hilligsøe}, \citenamefont {Paulsen}, \citenamefont {Thøgersen},
  \citenamefont {Keiding},\ and\ \citenamefont
  {Larsen}}]{hilligsoe_initial_2003}%
  \BibitemOpen
  \bibfield  {author} {\bibinfo {author} {\bibfnamefont {K.~M.}\ \bibnamefont
  {Hilligsøe}}, \bibinfo {author} {\bibfnamefont {H.~N.}\ \bibnamefont
  {Paulsen}}, \bibinfo {author} {\bibfnamefont {J.}~\bibnamefont {Thøgersen}},
  \bibinfo {author} {\bibfnamefont {S.~R.}\ \bibnamefont {Keiding}},\ and\
  \bibinfo {author} {\bibfnamefont {J.~J.}\ \bibnamefont {Larsen}},\ }\href
  {https://doi.org/10.1364/JOSAB.20.001887} {\bibfield  {journal} {\bibinfo
  {journal} {JOSA B}\ }\textbf {\bibinfo {volume} {20}},\ \bibinfo {pages}
  {1887} (\bibinfo {year} {2003})}\BibitemShut {NoStop}%
\bibitem [{\citenamefont {Yeom}\ \emph {et~al.}(2008)\citenamefont {Yeom},
  \citenamefont {M\"agi}, \citenamefont {Lamont}, \citenamefont {Roelens},
  \citenamefont {Fu},\ and\ \citenamefont
  {Eggleton}}]{yeom_low-threshold_2008}%
  \BibitemOpen
  \bibfield  {author} {\bibinfo {author} {\bibfnamefont {D.-I.}\ \bibnamefont
  {Yeom}}, \bibinfo {author} {\bibfnamefont {E.~C.}\ \bibnamefont {M\"agi}},
  \bibinfo {author} {\bibfnamefont {M.~R.~E.}\ \bibnamefont {Lamont}}, \bibinfo
  {author} {\bibfnamefont {M.~A.~F.}\ \bibnamefont {Roelens}}, \bibinfo
  {author} {\bibfnamefont {L.}~\bibnamefont {Fu}},\ and\ \bibinfo {author}
  {\bibfnamefont {B.~J.}\ \bibnamefont {Eggleton}},\ }\href
  {https://doi.org/10.1364/OL.33.000660} {\bibfield  {journal} {\bibinfo
  {journal} {Optics Letters}\ }\textbf {\bibinfo {volume} {33}},\ \bibinfo
  {pages} {660} (\bibinfo {year} {2008})}\BibitemShut {NoStop}%
\bibitem [{\citenamefont {Halir}\ \emph {et~al.}(2012)\citenamefont {Halir},
  \citenamefont {Okawachi}, \citenamefont {Levy}, \citenamefont {Foster},
  \citenamefont {Lipson},\ and\ \citenamefont
  {Gaeta}}]{halir_ultrabroadband_2012}%
  \BibitemOpen
  \bibfield  {author} {\bibinfo {author} {\bibfnamefont {R.}~\bibnamefont
  {Halir}}, \bibinfo {author} {\bibfnamefont {Y.}~\bibnamefont {Okawachi}},
  \bibinfo {author} {\bibfnamefont {J.~S.}\ \bibnamefont {Levy}}, \bibinfo
  {author} {\bibfnamefont {M.~A.}\ \bibnamefont {Foster}}, \bibinfo {author}
  {\bibfnamefont {M.}~\bibnamefont {Lipson}},\ and\ \bibinfo {author}
  {\bibfnamefont {A.~L.}\ \bibnamefont {Gaeta}},\ }\href
  {https://doi.org/10.1364/OL.37.001685} {\bibfield  {journal} {\bibinfo
  {journal} {Optics Letters}\ }\textbf {\bibinfo {volume} {37}},\ \bibinfo
  {pages} {1685} (\bibinfo {year} {2012})}\BibitemShut {NoStop}%
\bibitem [{\citenamefont {Leo}\ \emph {et~al.}(2014)\citenamefont {Leo},
  \citenamefont {Gorza}, \citenamefont {Safioui}, \citenamefont {Kockaert},
  \citenamefont {Coen}, \citenamefont {Dave}, \citenamefont {Kuyken},\ and\
  \citenamefont {Roelkens}}]{leo_dispersive_2014}%
  \BibitemOpen
  \bibfield  {author} {\bibinfo {author} {\bibfnamefont {F.}~\bibnamefont
  {Leo}}, \bibinfo {author} {\bibfnamefont {S.-P.}\ \bibnamefont {Gorza}},
  \bibinfo {author} {\bibfnamefont {J.}~\bibnamefont {Safioui}}, \bibinfo
  {author} {\bibfnamefont {P.}~\bibnamefont {Kockaert}}, \bibinfo {author}
  {\bibfnamefont {S.}~\bibnamefont {Coen}}, \bibinfo {author} {\bibfnamefont
  {U.}~\bibnamefont {Dave}}, \bibinfo {author} {\bibfnamefont {B.}~\bibnamefont
  {Kuyken}},\ and\ \bibinfo {author} {\bibfnamefont {G.}~\bibnamefont
  {Roelkens}},\ }\href {https://doi.org/10.1364/OL.39.003623} {\bibfield
  {journal} {\bibinfo  {journal} {Optics Letters}\ }\textbf {\bibinfo {volume}
  {39}},\ \bibinfo {pages} {3623} (\bibinfo {year} {2014})}\BibitemShut
  {NoStop}%
\bibitem [{\citenamefont {Guo}\ \emph {et~al.}(2018)\citenamefont {Guo},
  \citenamefont {Herkommer}, \citenamefont {Billat}, \citenamefont {Grassani},
  \citenamefont {Zhang}, \citenamefont {Pfeiffer}, \citenamefont {Weng},
  \citenamefont {Br\`es},\ and\ \citenamefont
  {Kippenberg}}]{guo_mid-infrared_2018}%
  \BibitemOpen
  \bibfield  {author} {\bibinfo {author} {\bibfnamefont {H.}~\bibnamefont
  {Guo}}, \bibinfo {author} {\bibfnamefont {C.}~\bibnamefont {Herkommer}},
  \bibinfo {author} {\bibfnamefont {A.}~\bibnamefont {Billat}}, \bibinfo
  {author} {\bibfnamefont {D.}~\bibnamefont {Grassani}}, \bibinfo {author}
  {\bibfnamefont {C.}~\bibnamefont {Zhang}}, \bibinfo {author} {\bibfnamefont
  {M.~H.~P.}\ \bibnamefont {Pfeiffer}}, \bibinfo {author} {\bibfnamefont
  {W.}~\bibnamefont {Weng}}, \bibinfo {author} {\bibfnamefont {C.-S.}\
  \bibnamefont {Br\`es}},\ and\ \bibinfo {author} {\bibfnamefont {T.~J.}\
  \bibnamefont {Kippenberg}},\ }\href
  {https://doi.org/10.1038/s41566-018-0144-1} {\bibfield  {journal} {\bibinfo
  {journal} {Nature Photonics}\ ,\ \bibinfo {pages} {1}} (\bibinfo {year}
  {2018})}\BibitemShut {NoStop}%
\bibitem [{\citenamefont {Wu}\ \emph {et~al.}(2013)\citenamefont {Wu},
  \citenamefont {Torres-Company}, \citenamefont {Leaird},\ and\ \citenamefont
  {Weiner}}]{wu_supercontinuum-based_2013}%
  \BibitemOpen
  \bibfield  {author} {\bibinfo {author} {\bibfnamefont {R.}~\bibnamefont
  {Wu}}, \bibinfo {author} {\bibfnamefont {V.}~\bibnamefont {Torres-Company}},
  \bibinfo {author} {\bibfnamefont {D.~E.}\ \bibnamefont {Leaird}},\ and\
  \bibinfo {author} {\bibfnamefont {A.~M.}\ \bibnamefont {Weiner}},\ }\href
  {https://doi.org/10.1364/OE.21.006045} {\bibfield  {journal} {\bibinfo
  {journal} {Optics Express}\ }\textbf {\bibinfo {volume} {21}},\ \bibinfo
  {pages} {6045} (\bibinfo {year} {2013})}\BibitemShut {NoStop}%
\bibitem [{\citenamefont {Beha}\ \emph {et~al.}(2017)\citenamefont {Beha},
  \citenamefont {Cole}, \citenamefont {Del’Haye}, \citenamefont {Coillet},
  \citenamefont {Diddams},\ and\ \citenamefont {Papp}}]{beha_electronic_2017}%
  \BibitemOpen
  \bibfield  {author} {\bibinfo {author} {\bibfnamefont {K.}~\bibnamefont
  {Beha}}, \bibinfo {author} {\bibfnamefont {D.~C.}\ \bibnamefont {Cole}},
  \bibinfo {author} {\bibfnamefont {P.}~\bibnamefont {Del’Haye}}, \bibinfo
  {author} {\bibfnamefont {A.}~\bibnamefont {Coillet}}, \bibinfo {author}
  {\bibfnamefont {S.~A.}\ \bibnamefont {Diddams}},\ and\ \bibinfo {author}
  {\bibfnamefont {S.~B.}\ \bibnamefont {Papp}},\ }\href
  {https://doi.org/10.1364/OPTICA.4.000406} {\bibfield  {journal} {\bibinfo
  {journal} {Optica}\ }\textbf {\bibinfo {volume} {4}},\ \bibinfo {pages} {406}
  (\bibinfo {year} {2017})}\BibitemShut {NoStop}%
\bibitem [{\citenamefont {Obrzud}\ \emph {et~al.}(2018)\citenamefont {Obrzud},
  \citenamefont {Rainer}, \citenamefont {Harutyunyan}, \citenamefont
  {Chazelas}, \citenamefont {Cecconi}, \citenamefont {Ghedina}, \citenamefont
  {Molinari}, \citenamefont {Kundermann}, \citenamefont {Lecomte},
  \citenamefont {Pepe}, \citenamefont {Wildi}, \citenamefont {Bouchy},\ and\
  \citenamefont {Herr}}]{obrzud_broadband_2018-1}%
  \BibitemOpen
  \bibfield  {author} {\bibinfo {author} {\bibfnamefont {E.}~\bibnamefont
  {Obrzud}}, \bibinfo {author} {\bibfnamefont {M.}~\bibnamefont {Rainer}},
  \bibinfo {author} {\bibfnamefont {A.}~\bibnamefont {Harutyunyan}}, \bibinfo
  {author} {\bibfnamefont {B.}~\bibnamefont {Chazelas}}, \bibinfo {author}
  {\bibfnamefont {M.}~\bibnamefont {Cecconi}}, \bibinfo {author} {\bibfnamefont
  {A.}~\bibnamefont {Ghedina}}, \bibinfo {author} {\bibfnamefont
  {E.}~\bibnamefont {Molinari}}, \bibinfo {author} {\bibfnamefont
  {S.}~\bibnamefont {Kundermann}}, \bibinfo {author} {\bibfnamefont
  {S.}~\bibnamefont {Lecomte}}, \bibinfo {author} {\bibfnamefont
  {F.}~\bibnamefont {Pepe}}, \bibinfo {author} {\bibfnamefont {F.}~\bibnamefont
  {Wildi}}, \bibinfo {author} {\bibfnamefont {F.}~\bibnamefont {Bouchy}},\ and\
  \bibinfo {author} {\bibfnamefont {T.}~\bibnamefont {Herr}},\ }\href
  {https://doi.org/10.1364/OE.26.034830} {\bibfield  {journal} {\bibinfo
  {journal} {Optics Express}\ }\textbf {\bibinfo {volume} {26}},\ \bibinfo
  {pages} {34830} (\bibinfo {year} {2018})}\BibitemShut {NoStop}%
\bibitem [{\citenamefont {Carlson}\ \emph {et~al.}(2018)\citenamefont
  {Carlson}, \citenamefont {Hickstein}, \citenamefont {Zhang}, \citenamefont
  {Metcalf}, \citenamefont {Quinlan}, \citenamefont {Diddams},\ and\
  \citenamefont {Papp}}]{carlson_ultrafast_2018}%
  \BibitemOpen
  \bibfield  {author} {\bibinfo {author} {\bibfnamefont {D.~R.}\ \bibnamefont
  {Carlson}}, \bibinfo {author} {\bibfnamefont {D.~D.}\ \bibnamefont
  {Hickstein}}, \bibinfo {author} {\bibfnamefont {W.}~\bibnamefont {Zhang}},
  \bibinfo {author} {\bibfnamefont {A.~J.}\ \bibnamefont {Metcalf}}, \bibinfo
  {author} {\bibfnamefont {F.}~\bibnamefont {Quinlan}}, \bibinfo {author}
  {\bibfnamefont {S.~A.}\ \bibnamefont {Diddams}},\ and\ \bibinfo {author}
  {\bibfnamefont {S.~B.}\ \bibnamefont {Papp}},\ }\href
  {https://doi.org/10.1126/science.aat6451} {\bibfield  {journal} {\bibinfo
  {journal} {Science}\ }\textbf {\bibinfo {volume} {361}},\ \bibinfo {pages}
  {1358} (\bibinfo {year} {2018})}\BibitemShut {NoStop}%
\bibitem [{\citenamefont {Leo}\ \emph {et~al.}(2013)\citenamefont {Leo},
  \citenamefont {Gelens}, \citenamefont {Emplit}, \citenamefont {Haelterman},\
  and\ \citenamefont {Coen}}]{leo_dynamics_2013}%
  \BibitemOpen
  \bibfield  {author} {\bibinfo {author} {\bibfnamefont {F.}~\bibnamefont
  {Leo}}, \bibinfo {author} {\bibfnamefont {L.}~\bibnamefont {Gelens}},
  \bibinfo {author} {\bibfnamefont {P.}~\bibnamefont {Emplit}}, \bibinfo
  {author} {\bibfnamefont {M.}~\bibnamefont {Haelterman}},\ and\ \bibinfo
  {author} {\bibfnamefont {S.}~\bibnamefont {Coen}},\ }\href
  {https://doi.org/10.1364/OE.21.009180} {\bibfield  {journal} {\bibinfo
  {journal} {Optics Express}\ }\textbf {\bibinfo {volume} {21}},\ \bibinfo
  {pages} {9180} (\bibinfo {year} {2013})}\BibitemShut {NoStop}%
\bibitem [{\citenamefont {Lucas}\ \emph
  {et~al.}(2017{\natexlab{a}})\citenamefont {Lucas}, \citenamefont {Karpov},
  \citenamefont {Guo}, \citenamefont {Gorodetsky},\ and\ \citenamefont
  {Kippenberg}}]{lucas_breathing_2017}%
  \BibitemOpen
  \bibfield  {author} {\bibinfo {author} {\bibfnamefont {E.}~\bibnamefont
  {Lucas}}, \bibinfo {author} {\bibfnamefont {M.}~\bibnamefont {Karpov}},
  \bibinfo {author} {\bibfnamefont {H.}~\bibnamefont {Guo}}, \bibinfo {author}
  {\bibfnamefont {M.~L.}\ \bibnamefont {Gorodetsky}},\ and\ \bibinfo {author}
  {\bibfnamefont {T.~J.}\ \bibnamefont {Kippenberg}},\ }\href
  {https://doi.org/10.1038/s41467-017-00719-w} {\bibfield  {journal} {\bibinfo
  {journal} {Nature Communications}\ }\textbf {\bibinfo {volume} {8}},\
  \bibinfo {pages} {736} (\bibinfo {year} {2017}{\natexlab{a}})}\BibitemShut
  {NoStop}%
\bibitem [{\citenamefont {Anderson}\ \emph {et~al.}(2016)\citenamefont
  {Anderson}, \citenamefont {Leo}, \citenamefont {Coen}, \citenamefont
  {Erkintalo},\ and\ \citenamefont {Murdoch}}]{anderson_observations_2016}%
  \BibitemOpen
  \bibfield  {author} {\bibinfo {author} {\bibfnamefont {M.}~\bibnamefont
  {Anderson}}, \bibinfo {author} {\bibfnamefont {F.}~\bibnamefont {Leo}},
  \bibinfo {author} {\bibfnamefont {S.}~\bibnamefont {Coen}}, \bibinfo {author}
  {\bibfnamefont {M.}~\bibnamefont {Erkintalo}},\ and\ \bibinfo {author}
  {\bibfnamefont {S.~G.}\ \bibnamefont {Murdoch}},\ }\href
  {https://doi.org/10.1364/OPTICA.3.001071} {\bibfield  {journal} {\bibinfo
  {journal} {Optica}\ }\textbf {\bibinfo {volume} {3}},\ \bibinfo {pages}
  {1071} (\bibinfo {year} {2016})}\BibitemShut {NoStop}%
\bibitem [{\citenamefont {Wang}\ \emph {et~al.}(2017)\citenamefont {Wang},
  \citenamefont {Leo}, \citenamefont {Fatome}, \citenamefont {Erkintalo},
  \citenamefont {Murdoch},\ and\ \citenamefont {Coen}}]{wang_universal_2017}%
  \BibitemOpen
  \bibfield  {author} {\bibinfo {author} {\bibfnamefont {Y.}~\bibnamefont
  {Wang}}, \bibinfo {author} {\bibfnamefont {F.}~\bibnamefont {Leo}}, \bibinfo
  {author} {\bibfnamefont {J.}~\bibnamefont {Fatome}}, \bibinfo {author}
  {\bibfnamefont {M.}~\bibnamefont {Erkintalo}}, \bibinfo {author}
  {\bibfnamefont {S.~G.}\ \bibnamefont {Murdoch}},\ and\ \bibinfo {author}
  {\bibfnamefont {S.}~\bibnamefont {Coen}},\ }\href
  {https://doi.org/10.1364/OPTICA.4.000855} {\bibfield  {journal} {\bibinfo
  {journal} {Optica}\ }\textbf {\bibinfo {volume} {4}},\ \bibinfo {pages} {855}
  (\bibinfo {year} {2017})}\BibitemShut {NoStop}%
\bibitem [{\citenamefont {Akhmediev}\ and\ \citenamefont
  {Ankiewicz}(2008)}]{akhmediev_dissipative_2008}%
  \BibitemOpen
  \bibinfo {editor} {\bibfnamefont {N.}~\bibnamefont {Akhmediev}}\ and\
  \bibinfo {editor} {\bibfnamefont {A.}~\bibnamefont {Ankiewicz}},\ eds.,\
  \href {https://www.springer.com/gp/book/9783540782162} {\emph {\bibinfo
  {title} {Dissipative {Solitons}: {From} {Optics} to {Biology} and
  {Medicine}}}},\ Lecture {Notes} in {Physics}\ (\bibinfo  {publisher}
  {Springer-Verlag},\ \bibinfo {address} {Berlin Heidelberg},\ \bibinfo {year}
  {2008})\BibitemShut {NoStop}%
\bibitem [{\citenamefont {{Ewelina Obrzud}}\ \emph {et~al.}(2019)\citenamefont
  {{Ewelina Obrzud}}, \citenamefont {{Monica Rainer}}, \citenamefont {{Avet
  Harutyunyan}}, \citenamefont {{Miles H. Anderson}}, \citenamefont {{Junqiu
  Liu}}, \citenamefont {{Michael Geiselmann}}, \citenamefont {{Bruno
  Chazelas}}, \citenamefont {{Stefan Kundermann}}, \citenamefont {{Steve
  Lecomte}}, \citenamefont {{Massimo Cecconi}}, \citenamefont {{Adriano
  Ghedina}}, \citenamefont {{Emilio Molinari}}, \citenamefont {{Francesco
  Pepe}}, \citenamefont {{François Wildi}}, \citenamefont {{François
  Bouchy}}, \citenamefont {{Tobias J. Kippenberg}},\ and\ \citenamefont
  {{Tobias Herr}}}]{ewelina_obrzud_microphotonic_2019}%
  \BibitemOpen
  \bibfield  {author} {\bibinfo {author} {\bibnamefont {{Ewelina Obrzud}}},
  \bibinfo {author} {\bibnamefont {{Monica Rainer}}}, \bibinfo {author}
  {\bibnamefont {{Avet Harutyunyan}}}, \bibinfo {author} {\bibnamefont {{Miles
  H. Anderson}}}, \bibinfo {author} {\bibnamefont {{Junqiu Liu}}}, \bibinfo
  {author} {\bibnamefont {{Michael Geiselmann}}}, \bibinfo {author}
  {\bibnamefont {{Bruno Chazelas}}}, \bibinfo {author} {\bibnamefont {{Stefan
  Kundermann}}}, \bibinfo {author} {\bibnamefont {{Steve Lecomte}}}, \bibinfo
  {author} {\bibnamefont {{Massimo Cecconi}}}, \bibinfo {author} {\bibnamefont
  {{Adriano Ghedina}}}, \bibinfo {author} {\bibnamefont {{Emilio Molinari}}},
  \bibinfo {author} {\bibnamefont {{Francesco Pepe}}}, \bibinfo {author}
  {\bibnamefont {{François Wildi}}}, \bibinfo {author} {\bibnamefont
  {{François Bouchy}}}, \bibinfo {author} {\bibnamefont {{Tobias J.
  Kippenberg}}},\ and\ \bibinfo {author} {\bibnamefont {{Tobias Herr}}},\
  }\href {https://doi.org/10.1038/s41566-018-0309-y} {\bibfield  {journal}
  {\bibinfo  {journal} {Nature Photonics}\ }\textbf {\bibinfo {volume} {13}},\
  \bibinfo {pages} {31} (\bibinfo {year} {2019})}\BibitemShut {NoStop}%
\bibitem [{\citenamefont {Suh}\ \emph {et~al.}(2019)\citenamefont {Suh},
  \citenamefont {Yi}, \citenamefont {Lai}, \citenamefont {Leifer},
  \citenamefont {Grudinin}, \citenamefont {Vasisht}, \citenamefont {Martin},
  \citenamefont {Fitzgerald}, \citenamefont {Doppmann}, \citenamefont {Wang},
  \citenamefont {Mawet}, \citenamefont {Papp}, \citenamefont {Diddams},
  \citenamefont {Beichman},\ and\ \citenamefont {Vahala}}]{suh_searching_2019}%
  \BibitemOpen
  \bibfield  {author} {\bibinfo {author} {\bibfnamefont {M.-G.}\ \bibnamefont
  {Suh}}, \bibinfo {author} {\bibfnamefont {X.}~\bibnamefont {Yi}}, \bibinfo
  {author} {\bibfnamefont {Y.-H.}\ \bibnamefont {Lai}}, \bibinfo {author}
  {\bibfnamefont {S.}~\bibnamefont {Leifer}}, \bibinfo {author} {\bibfnamefont
  {I.~S.}\ \bibnamefont {Grudinin}}, \bibinfo {author} {\bibfnamefont
  {G.}~\bibnamefont {Vasisht}}, \bibinfo {author} {\bibfnamefont {E.~C.}\
  \bibnamefont {Martin}}, \bibinfo {author} {\bibfnamefont {M.~P.}\
  \bibnamefont {Fitzgerald}}, \bibinfo {author} {\bibfnamefont
  {G.}~\bibnamefont {Doppmann}}, \bibinfo {author} {\bibfnamefont
  {J.}~\bibnamefont {Wang}}, \bibinfo {author} {\bibfnamefont {D.}~\bibnamefont
  {Mawet}}, \bibinfo {author} {\bibfnamefont {S.~B.}\ \bibnamefont {Papp}},
  \bibinfo {author} {\bibfnamefont {S.~A.}\ \bibnamefont {Diddams}}, \bibinfo
  {author} {\bibfnamefont {C.}~\bibnamefont {Beichman}},\ and\ \bibinfo
  {author} {\bibfnamefont {K.}~\bibnamefont {Vahala}},\ }\href
  {https://doi.org/10.1038/s41566-018-0312-3} {\bibfield  {journal} {\bibinfo
  {journal} {Nature Photonics}\ }\textbf {\bibinfo {volume} {13}},\ \bibinfo
  {pages} {25} (\bibinfo {year} {2019})}\BibitemShut {NoStop}%
\bibitem [{\citenamefont {Hu}\ \emph {et~al.}(2018)\citenamefont {Hu},
  \citenamefont {Ros}, \citenamefont {Pu}, \citenamefont {Ye}, \citenamefont
  {Ingerslev}, \citenamefont {Silva}, \citenamefont {Nooruzzaman},
  \citenamefont {Amma}, \citenamefont {Sasaki}, \citenamefont {Mizuno},
  \citenamefont {Miyamoto}, \citenamefont {Ottaviano}, \citenamefont
  {Semenova}, \citenamefont {Guan}, \citenamefont {Zibar}, \citenamefont
  {Galili}, \citenamefont {Yvind}, \citenamefont {Morioka},\ and\ \citenamefont
  {Oxenløwe}}]{hu_single-source_2018}%
  \BibitemOpen
  \bibfield  {author} {\bibinfo {author} {\bibfnamefont {H.}~\bibnamefont
  {Hu}}, \bibinfo {author} {\bibfnamefont {F.~D.}\ \bibnamefont {Ros}},
  \bibinfo {author} {\bibfnamefont {M.}~\bibnamefont {Pu}}, \bibinfo {author}
  {\bibfnamefont {F.}~\bibnamefont {Ye}}, \bibinfo {author} {\bibfnamefont
  {K.}~\bibnamefont {Ingerslev}}, \bibinfo {author} {\bibfnamefont {E.~P.~d.}\
  \bibnamefont {Silva}}, \bibinfo {author} {\bibfnamefont {M.}~\bibnamefont
  {Nooruzzaman}}, \bibinfo {author} {\bibfnamefont {Y.}~\bibnamefont {Amma}},
  \bibinfo {author} {\bibfnamefont {Y.}~\bibnamefont {Sasaki}}, \bibinfo
  {author} {\bibfnamefont {T.}~\bibnamefont {Mizuno}}, \bibinfo {author}
  {\bibfnamefont {Y.}~\bibnamefont {Miyamoto}}, \bibinfo {author}
  {\bibfnamefont {L.}~\bibnamefont {Ottaviano}}, \bibinfo {author}
  {\bibfnamefont {E.}~\bibnamefont {Semenova}}, \bibinfo {author}
  {\bibfnamefont {P.}~\bibnamefont {Guan}}, \bibinfo {author} {\bibfnamefont
  {D.}~\bibnamefont {Zibar}}, \bibinfo {author} {\bibfnamefont
  {M.}~\bibnamefont {Galili}}, \bibinfo {author} {\bibfnamefont
  {K.}~\bibnamefont {Yvind}}, \bibinfo {author} {\bibfnamefont
  {T.}~\bibnamefont {Morioka}},\ and\ \bibinfo {author} {\bibfnamefont {L.~K.}\
  \bibnamefont {Oxenløwe}},\ }\href
  {https://doi.org/10.1038/s41566-018-0205-5} {\bibfield  {journal} {\bibinfo
  {journal} {Nature Photonics}\ ,\ \bibinfo {pages} {1}} (\bibinfo {year}
  {2018})}\BibitemShut {NoStop}%
\bibitem [{\citenamefont {Malinowski}\ \emph {et~al.}(2017)\citenamefont
  {Malinowski}, \citenamefont {Rao}, \citenamefont {Delfyett},\ and\
  \citenamefont {Fathpour}}]{malinowski_optical_2017}%
  \BibitemOpen
  \bibfield  {author} {\bibinfo {author} {\bibfnamefont {M.}~\bibnamefont
  {Malinowski}}, \bibinfo {author} {\bibfnamefont {A.}~\bibnamefont {Rao}},
  \bibinfo {author} {\bibfnamefont {P.}~\bibnamefont {Delfyett}},\ and\
  \bibinfo {author} {\bibfnamefont {S.}~\bibnamefont {Fathpour}},\ }\href
  {https://doi.org/10.1063/1.4983113} {\bibfield  {journal} {\bibinfo
  {journal} {APL Photonics}\ }\textbf {\bibinfo {volume} {2}},\ \bibinfo
  {pages} {066101} (\bibinfo {year} {2017})}\BibitemShut {NoStop}%
\bibitem [{\citenamefont {Lilienfein}\ \emph {et~al.}(2019)\citenamefont
  {Lilienfein}, \citenamefont {Hofer}, \citenamefont {Högner}, \citenamefont
  {Saule}, \citenamefont {Trubetskov}, \citenamefont {Pervak}, \citenamefont
  {Fill}, \citenamefont {Riek}, \citenamefont {Leitenstorfer}, \citenamefont
  {Limpert}, \citenamefont {Krausz},\ and\ \citenamefont
  {Pupeza}}]{lilienfein_temporal_2019}%
  \BibitemOpen
  \bibfield  {author} {\bibinfo {author} {\bibfnamefont {N.}~\bibnamefont
  {Lilienfein}}, \bibinfo {author} {\bibfnamefont {C.}~\bibnamefont {Hofer}},
  \bibinfo {author} {\bibfnamefont {M.}~\bibnamefont {Högner}}, \bibinfo
  {author} {\bibfnamefont {T.}~\bibnamefont {Saule}}, \bibinfo {author}
  {\bibfnamefont {M.}~\bibnamefont {Trubetskov}}, \bibinfo {author}
  {\bibfnamefont {V.}~\bibnamefont {Pervak}}, \bibinfo {author} {\bibfnamefont
  {E.}~\bibnamefont {Fill}}, \bibinfo {author} {\bibfnamefont {C.}~\bibnamefont
  {Riek}}, \bibinfo {author} {\bibfnamefont {A.}~\bibnamefont {Leitenstorfer}},
  \bibinfo {author} {\bibfnamefont {J.}~\bibnamefont {Limpert}}, \bibinfo
  {author} {\bibfnamefont {F.}~\bibnamefont {Krausz}},\ and\ \bibinfo {author}
  {\bibfnamefont {I.}~\bibnamefont {Pupeza}},\ }\href
  {https://doi.org/10.1038/s41566-018-0341-y} {\bibfield  {journal} {\bibinfo
  {journal} {Nature Photonics}\ ,\ \bibinfo {pages} {1}} (\bibinfo {year}
  {2019})}\BibitemShut {NoStop}%
\bibitem [{\citenamefont {Okawachi}\ \emph {et~al.}(2014)\citenamefont
  {Okawachi}, \citenamefont {Lamont}, \citenamefont {Luke}, \citenamefont
  {Carvalho}, \citenamefont {Yu}, \citenamefont {Lipson},\ and\ \citenamefont
  {Gaeta}}]{okawachi_bandwidth_2014}%
  \BibitemOpen
  \bibfield  {author} {\bibinfo {author} {\bibfnamefont {Y.}~\bibnamefont
  {Okawachi}}, \bibinfo {author} {\bibfnamefont {M.~R.~E.}\ \bibnamefont
  {Lamont}}, \bibinfo {author} {\bibfnamefont {K.}~\bibnamefont {Luke}},
  \bibinfo {author} {\bibfnamefont {D.~O.}\ \bibnamefont {Carvalho}}, \bibinfo
  {author} {\bibfnamefont {M.}~\bibnamefont {Yu}}, \bibinfo {author}
  {\bibfnamefont {M.}~\bibnamefont {Lipson}},\ and\ \bibinfo {author}
  {\bibfnamefont {A.~L.}\ \bibnamefont {Gaeta}},\ }\href
  {https://doi.org/10.1364/OL.39.003535} {\bibfield  {journal} {\bibinfo
  {journal} {Optics Letters}\ }\textbf {\bibinfo {volume} {39}},\ \bibinfo
  {pages} {3535} (\bibinfo {year} {2014})}\BibitemShut {NoStop}%
\bibitem [{\citenamefont {Okawachi}\ \emph {et~al.}(2018)\citenamefont
  {Okawachi}, \citenamefont {Yu}, \citenamefont {Cardenas}, \citenamefont {Ji},
  \citenamefont {Klenner}, \citenamefont {Lipson},\ and\ \citenamefont
  {Gaeta}}]{okawachi_carrier_2018}%
  \BibitemOpen
  \bibfield  {author} {\bibinfo {author} {\bibfnamefont {Y.}~\bibnamefont
  {Okawachi}}, \bibinfo {author} {\bibfnamefont {M.}~\bibnamefont {Yu}},
  \bibinfo {author} {\bibfnamefont {J.}~\bibnamefont {Cardenas}}, \bibinfo
  {author} {\bibfnamefont {X.}~\bibnamefont {Ji}}, \bibinfo {author}
  {\bibfnamefont {A.}~\bibnamefont {Klenner}}, \bibinfo {author} {\bibfnamefont
  {M.}~\bibnamefont {Lipson}},\ and\ \bibinfo {author} {\bibfnamefont {A.~L.}\
  \bibnamefont {Gaeta}},\ }\href {https://doi.org/10.1364/OL.43.004627}
  {\bibfield  {journal} {\bibinfo  {journal} {Optics Letters}\ }\textbf
  {\bibinfo {volume} {43}},\ \bibinfo {pages} {4627} (\bibinfo {year}
  {2018})}\BibitemShut {NoStop}%
\bibitem [{\citenamefont {Hendry}\ \emph {et~al.}(2018)\citenamefont {Hendry},
  \citenamefont {Chen}, \citenamefont {Wang}, \citenamefont {Garbin},
  \citenamefont {Javaloyes}, \citenamefont {Oppo}, \citenamefont {Coen},
  \citenamefont {Murdoch},\ and\ \citenamefont
  {Erkintalo}}]{hendry_spontaneous_2018}%
  \BibitemOpen
  \bibfield  {author} {\bibinfo {author} {\bibfnamefont {I.}~\bibnamefont
  {Hendry}}, \bibinfo {author} {\bibfnamefont {W.}~\bibnamefont {Chen}},
  \bibinfo {author} {\bibfnamefont {Y.}~\bibnamefont {Wang}}, \bibinfo {author}
  {\bibfnamefont {B.}~\bibnamefont {Garbin}}, \bibinfo {author} {\bibfnamefont
  {J.}~\bibnamefont {Javaloyes}}, \bibinfo {author} {\bibfnamefont {G.-L.}\
  \bibnamefont {Oppo}}, \bibinfo {author} {\bibfnamefont {S.}~\bibnamefont
  {Coen}}, \bibinfo {author} {\bibfnamefont {S.~G.}\ \bibnamefont {Murdoch}},\
  and\ \bibinfo {author} {\bibfnamefont {M.}~\bibnamefont {Erkintalo}},\
  }\bibfield  {journal} {\bibinfo  {journal} {Physical Review A}\ }\textbf
  {\bibinfo {volume} {97}},\ \href {https://doi.org/10.1103/PhysRevA.97.053834}
  {10.1103/PhysRevA.97.053834} (\bibinfo {year} {2018})\BibitemShut {NoStop}%
\bibitem [{\citenamefont {Ishizawa}\ \emph {et~al.}(2013)\citenamefont
  {Ishizawa}, \citenamefont {Nishikawa}, \citenamefont {Mizutori},
  \citenamefont {Takara}, \citenamefont {Takada}, \citenamefont {Sogawa},\ and\
  \citenamefont {Koga}}]{ishizawa_phase-noise_2013}%
  \BibitemOpen
  \bibfield  {author} {\bibinfo {author} {\bibfnamefont {A.}~\bibnamefont
  {Ishizawa}}, \bibinfo {author} {\bibfnamefont {T.}~\bibnamefont {Nishikawa}},
  \bibinfo {author} {\bibfnamefont {A.}~\bibnamefont {Mizutori}}, \bibinfo
  {author} {\bibfnamefont {H.}~\bibnamefont {Takara}}, \bibinfo {author}
  {\bibfnamefont {A.}~\bibnamefont {Takada}}, \bibinfo {author} {\bibfnamefont
  {T.}~\bibnamefont {Sogawa}},\ and\ \bibinfo {author} {\bibfnamefont
  {M.}~\bibnamefont {Koga}},\ }\href {https://doi.org/10.1364/OE.21.029186}
  {\bibfield  {journal} {\bibinfo  {journal} {Optics Express}\ }\textbf
  {\bibinfo {volume} {21}},\ \bibinfo {pages} {29186} (\bibinfo {year}
  {2013})}\BibitemShut {NoStop}%
\bibitem [{\citenamefont {Kobayashi}\ \emph {et~al.}(1988)\citenamefont
  {Kobayashi}, \citenamefont {Yao}, \citenamefont {Amano}, \citenamefont
  {Fukushima}, \citenamefont {Morimoto},\ and\ \citenamefont
  {Sueta}}]{kobayashi_optical_1988}%
  \BibitemOpen
  \bibfield  {author} {\bibinfo {author} {\bibfnamefont {T.}~\bibnamefont
  {Kobayashi}}, \bibinfo {author} {\bibfnamefont {H.}~\bibnamefont {Yao}},
  \bibinfo {author} {\bibfnamefont {K.}~\bibnamefont {Amano}}, \bibinfo
  {author} {\bibfnamefont {Y.}~\bibnamefont {Fukushima}}, \bibinfo {author}
  {\bibfnamefont {A.}~\bibnamefont {Morimoto}},\ and\ \bibinfo {author}
  {\bibfnamefont {T.}~\bibnamefont {Sueta}},\ }\href
  {https://doi.org/10.1109/3.135} {\bibfield  {journal} {\bibinfo  {journal}
  {IEEE Journal of Quantum Electronics}\ }\textbf {\bibinfo {volume} {24}},\
  \bibinfo {pages} {382} (\bibinfo {year} {1988})}\BibitemShut {NoStop}%
\bibitem [{\citenamefont {Fujiwara}\ \emph {et~al.}(2003)\citenamefont
  {Fujiwara}, \citenamefont {Teshima}, \citenamefont {Kani}, \citenamefont
  {Suzuki}, \citenamefont {Takachio},\ and\ \citenamefont
  {Iwatsuki}}]{fujiwara_optical_2003}%
  \BibitemOpen
  \bibfield  {author} {\bibinfo {author} {\bibfnamefont {M.}~\bibnamefont
  {Fujiwara}}, \bibinfo {author} {\bibfnamefont {M.}~\bibnamefont {Teshima}},
  \bibinfo {author} {\bibfnamefont {J.}~\bibnamefont {Kani}}, \bibinfo {author}
  {\bibfnamefont {H.}~\bibnamefont {Suzuki}}, \bibinfo {author} {\bibfnamefont
  {N.}~\bibnamefont {Takachio}},\ and\ \bibinfo {author} {\bibfnamefont
  {K.}~\bibnamefont {Iwatsuki}},\ }\href
  {https://doi.org/10.1109/JLT.2003.819147} {\bibfield  {journal} {\bibinfo
  {journal} {Journal of Lightwave Technology}\ }\textbf {\bibinfo {volume}
  {21}},\ \bibinfo {pages} {2705} (\bibinfo {year} {2003})}\BibitemShut
  {NoStop}%
\bibitem [{\citenamefont {Coen}\ and\ \citenamefont
  {Erkintalo}(2013)}]{coen_universal_2013}%
  \BibitemOpen
  \bibfield  {author} {\bibinfo {author} {\bibfnamefont {S.}~\bibnamefont
  {Coen}}\ and\ \bibinfo {author} {\bibfnamefont {M.}~\bibnamefont
  {Erkintalo}},\ }\href {https://doi.org/10.1364/OL.38.001790} {\bibfield
  {journal} {\bibinfo  {journal} {Optics Letters}\ }\textbf {\bibinfo {volume}
  {38}},\ \bibinfo {pages} {1790} (\bibinfo {year} {2013})}\BibitemShut
  {NoStop}%
\bibitem [{\citenamefont {Lucas}\ \emph
  {et~al.}(2017{\natexlab{b}})\citenamefont {Lucas}, \citenamefont {Guo},
  \citenamefont {Jost}, \citenamefont {Karpov},\ and\ \citenamefont
  {Kippenberg}}]{lucas_detuning-dependent_2017}%
  \BibitemOpen
  \bibfield  {author} {\bibinfo {author} {\bibfnamefont {E.}~\bibnamefont
  {Lucas}}, \bibinfo {author} {\bibfnamefont {H.}~\bibnamefont {Guo}}, \bibinfo
  {author} {\bibfnamefont {J.~D.}\ \bibnamefont {Jost}}, \bibinfo {author}
  {\bibfnamefont {M.}~\bibnamefont {Karpov}},\ and\ \bibinfo {author}
  {\bibfnamefont {T.~J.}\ \bibnamefont {Kippenberg}},\ }\href
  {https://doi.org/10.1103/PhysRevA.95.043822} {\bibfield  {journal} {\bibinfo
  {journal} {Physical Review A}\ }\textbf {\bibinfo {volume} {95}},\ \bibinfo
  {pages} {043822} (\bibinfo {year} {2017}{\natexlab{b}})}\BibitemShut
  {NoStop}%
\bibitem [{\citenamefont {Karpov}\ \emph {et~al.}(2016)\citenamefont {Karpov},
  \citenamefont {Guo}, \citenamefont {Kordts}, \citenamefont {Brasch},
  \citenamefont {Pfeiffer}, \citenamefont {Zervas}, \citenamefont
  {Geiselmann},\ and\ \citenamefont {Kippenberg}}]{karpov_raman_2016}%
  \BibitemOpen
  \bibfield  {author} {\bibinfo {author} {\bibfnamefont {M.}~\bibnamefont
  {Karpov}}, \bibinfo {author} {\bibfnamefont {H.}~\bibnamefont {Guo}},
  \bibinfo {author} {\bibfnamefont {A.}~\bibnamefont {Kordts}}, \bibinfo
  {author} {\bibfnamefont {V.}~\bibnamefont {Brasch}}, \bibinfo {author}
  {\bibfnamefont {M.~H.}\ \bibnamefont {Pfeiffer}}, \bibinfo {author}
  {\bibfnamefont {M.}~\bibnamefont {Zervas}}, \bibinfo {author} {\bibfnamefont
  {M.}~\bibnamefont {Geiselmann}},\ and\ \bibinfo {author} {\bibfnamefont
  {T.~J.}\ \bibnamefont {Kippenberg}},\ }\href
  {https://doi.org/10.1103/PhysRevLett.116.103902} {\bibfield  {journal}
  {\bibinfo  {journal} {Physical Review Letters}\ }\textbf {\bibinfo {volume}
  {116}},\ \bibinfo {pages} {103902} (\bibinfo {year} {2016})}\BibitemShut
  {NoStop}%
\bibitem [{\citenamefont {Yi}\ \emph {et~al.}(2016)\citenamefont {Yi},
  \citenamefont {Yang}, \citenamefont {Yang},\ and\ \citenamefont
  {Vahala}}]{yi_theory_2016}%
  \BibitemOpen
  \bibfield  {author} {\bibinfo {author} {\bibfnamefont {X.}~\bibnamefont
  {Yi}}, \bibinfo {author} {\bibfnamefont {Q.-F.}\ \bibnamefont {Yang}},
  \bibinfo {author} {\bibfnamefont {K.~Y.}\ \bibnamefont {Yang}},\ and\
  \bibinfo {author} {\bibfnamefont {K.}~\bibnamefont {Vahala}},\ }\href
  {https://doi.org/10.1364/OL.41.003419} {\bibfield  {journal} {\bibinfo
  {journal} {Optics Letters}\ }\textbf {\bibinfo {volume} {41}},\ \bibinfo
  {pages} {3419} (\bibinfo {year} {2016})}\BibitemShut {NoStop}%
\bibitem [{\citenamefont {Lamb}\ \emph {et~al.}(2018)\citenamefont {Lamb},
  \citenamefont {Carlson}, \citenamefont {Hickstein}, \citenamefont {Stone},
  \citenamefont {Diddams},\ and\ \citenamefont
  {Papp}}]{lamb_optical-frequency_2018}%
  \BibitemOpen
  \bibfield  {author} {\bibinfo {author} {\bibfnamefont {E.~S.}\ \bibnamefont
  {Lamb}}, \bibinfo {author} {\bibfnamefont {D.~R.}\ \bibnamefont {Carlson}},
  \bibinfo {author} {\bibfnamefont {D.~D.}\ \bibnamefont {Hickstein}}, \bibinfo
  {author} {\bibfnamefont {J.~R.}\ \bibnamefont {Stone}}, \bibinfo {author}
  {\bibfnamefont {S.~A.}\ \bibnamefont {Diddams}},\ and\ \bibinfo {author}
  {\bibfnamefont {S.~B.}\ \bibnamefont {Papp}},\ }\href
  {https://doi.org/10.1103/PhysRevApplied.9.024030} {\bibfield  {journal}
  {\bibinfo  {journal} {Physical Review Applied}\ }\textbf {\bibinfo {volume}
  {9}},\ \bibinfo {pages} {024030} (\bibinfo {year} {2018})}\BibitemShut
  {NoStop}%
\bibitem [{\citenamefont {Brasch}\ \emph {et~al.}(2016)\citenamefont {Brasch},
  \citenamefont {Geiselmann}, \citenamefont {Pfeiffer},\ and\ \citenamefont
  {Kippenberg}}]{brasch_bringing_2016}%
  \BibitemOpen
  \bibfield  {author} {\bibinfo {author} {\bibfnamefont {V.}~\bibnamefont
  {Brasch}}, \bibinfo {author} {\bibfnamefont {M.}~\bibnamefont {Geiselmann}},
  \bibinfo {author} {\bibfnamefont {M.~H.~P.}\ \bibnamefont {Pfeiffer}},\ and\
  \bibinfo {author} {\bibfnamefont {T.~J.}\ \bibnamefont {Kippenberg}},\ }\href
  {https://doi.org/10.1364/OE.24.029312} {\bibfield  {journal} {\bibinfo
  {journal} {Optics Express}\ }\textbf {\bibinfo {volume} {24}},\ \bibinfo
  {pages} {29312} (\bibinfo {year} {2016})}\BibitemShut {NoStop}%
\bibitem [{\citenamefont {Liu}\ \emph {et~al.}(2019)\citenamefont {Liu},
  \citenamefont {Lucas}, \citenamefont {Raja}, \citenamefont {He},
  \citenamefont {Riemensberger}, \citenamefont {Wang}, \citenamefont {Karpov},
  \citenamefont {Guo}, \citenamefont {Bouchand},\ and\ \citenamefont
  {Kippenberg}}]{liu_nanophotonic_2019}%
  \BibitemOpen
  \bibfield  {author} {\bibinfo {author} {\bibfnamefont {J.}~\bibnamefont
  {Liu}}, \bibinfo {author} {\bibfnamefont {E.}~\bibnamefont {Lucas}}, \bibinfo
  {author} {\bibfnamefont {A.~S.}\ \bibnamefont {Raja}}, \bibinfo {author}
  {\bibfnamefont {J.}~\bibnamefont {He}}, \bibinfo {author} {\bibfnamefont
  {J.}~\bibnamefont {Riemensberger}}, \bibinfo {author} {\bibfnamefont {R.~N.}\
  \bibnamefont {Wang}}, \bibinfo {author} {\bibfnamefont {M.}~\bibnamefont
  {Karpov}}, \bibinfo {author} {\bibfnamefont {H.}~\bibnamefont {Guo}},
  \bibinfo {author} {\bibfnamefont {R.}~\bibnamefont {Bouchand}},\ and\
  \bibinfo {author} {\bibfnamefont {T.~J.}\ \bibnamefont {Kippenberg}},\ }\href
  {http://arxiv.org/abs/1901.10372} {\bibfield  {journal} {\bibinfo  {journal}
  {arXiv:1901.10372 [physics]}\ } (\bibinfo {year} {2019})},\ \bibinfo {note}
  {arXiv: 1901.10372}\BibitemShut {NoStop}%
\bibitem [{\citenamefont {Lugiato}\ and\ \citenamefont
  {Lefever}(1987)}]{lugiato_spatial_1987}%
  \BibitemOpen
  \bibfield  {author} {\bibinfo {author} {\bibfnamefont {L.~A.}\ \bibnamefont
  {Lugiato}}\ and\ \bibinfo {author} {\bibfnamefont {R.}~\bibnamefont
  {Lefever}},\ }\href {https://doi.org/10.1103/PhysRevLett.58.2209} {\bibfield
  {journal} {\bibinfo  {journal} {Physical Review Letters}\ }\textbf {\bibinfo
  {volume} {58}},\ \bibinfo {pages} {2209} (\bibinfo {year}
  {1987})}\BibitemShut {NoStop}%
\bibitem [{\citenamefont {Agrawal}(1995)}]{AGRAWAL199560}%
  \BibitemOpen
  \bibfield  {author} {\bibinfo {author} {\bibfnamefont {G.~P.}\ \bibnamefont
  {Agrawal}},\ }in\ \href
  {https://doi.org/https://doi.org/10.1016/B978-0-12-045142-5.50009-X} {\emph
  {\bibinfo {booktitle} {Nonlinear Fiber Optics (Second Edition)}}},\ \bibinfo
  {series and number} {Optics and Photonics},\ \bibinfo {editor} {edited by\
  \bibinfo {editor} {\bibfnamefont {G.~P.}\ \bibnamefont {Agrawal}}}\ (\bibinfo
   {publisher} {Academic Press},\ \bibinfo {address} {Boston},\ \bibinfo {year}
  {1995})\ \bibinfo {edition} {second edition}\ ed.,\ pp.\ \bibinfo {pages} {60
  -- 88}\BibitemShut {NoStop}%
\bibitem [{\citenamefont {Guo}\ \emph {et~al.}(2017)\citenamefont {Guo},
  \citenamefont {Karpov}, \citenamefont {Lucas}, \citenamefont {Kordts},
  \citenamefont {Pfeiffer}, \citenamefont {Brasch}, \citenamefont {Lihachev},
  \citenamefont {Lobanov}, \citenamefont {Gorodetsky},\ and\ \citenamefont
  {Kippenberg}}]{guo_universal_2017}%
  \BibitemOpen
  \bibfield  {author} {\bibinfo {author} {\bibfnamefont {H.}~\bibnamefont
  {Guo}}, \bibinfo {author} {\bibfnamefont {M.}~\bibnamefont {Karpov}},
  \bibinfo {author} {\bibfnamefont {E.}~\bibnamefont {Lucas}}, \bibinfo
  {author} {\bibfnamefont {A.}~\bibnamefont {Kordts}}, \bibinfo {author}
  {\bibfnamefont {M.~H.~P.}\ \bibnamefont {Pfeiffer}}, \bibinfo {author}
  {\bibfnamefont {V.}~\bibnamefont {Brasch}}, \bibinfo {author} {\bibfnamefont
  {G.}~\bibnamefont {Lihachev}}, \bibinfo {author} {\bibfnamefont {V.~E.}\
  \bibnamefont {Lobanov}}, \bibinfo {author} {\bibfnamefont {M.~L.}\
  \bibnamefont {Gorodetsky}},\ and\ \bibinfo {author} {\bibfnamefont {T.~J.}\
  \bibnamefont {Kippenberg}},\ }\href {https://doi.org/10.1038/nphys3893}
  {\bibfield  {journal} {\bibinfo  {journal} {Nature Physics}\ }\textbf
  {\bibinfo {volume} {13}},\ \bibinfo {pages} {94} (\bibinfo {year}
  {2017})}\BibitemShut {NoStop}%
\bibitem [{\citenamefont {Domenico}\ \emph {et~al.}(2010)\citenamefont
  {Domenico}, \citenamefont {Schilt},\ and\ \citenamefont
  {Thomann}}]{domenico_simple_2010}%
  \BibitemOpen
  \bibfield  {author} {\bibinfo {author} {\bibfnamefont {G.~D.}\ \bibnamefont
  {Domenico}}, \bibinfo {author} {\bibfnamefont {S.}~\bibnamefont {Schilt}},\
  and\ \bibinfo {author} {\bibfnamefont {P.}~\bibnamefont {Thomann}},\ }\href
  {https://doi.org/10.1364/AO.49.004801} {\bibfield  {journal} {\bibinfo
  {journal} {Applied Optics}\ }\textbf {\bibinfo {volume} {49}},\ \bibinfo
  {pages} {4801} (\bibinfo {year} {2010})}\BibitemShut {NoStop}%
\bibitem [{\citenamefont {Schiemangk}\ \emph {et~al.}(2014)\citenamefont
  {Schiemangk}, \citenamefont {Spießberger}, \citenamefont {Wicht},
  \citenamefont {Erbert}, \citenamefont {Tr\"ankle},\ and\ \citenamefont
  {Peters}}]{schiemangk_accurate_2014}%
  \BibitemOpen
  \bibfield  {author} {\bibinfo {author} {\bibfnamefont {M.}~\bibnamefont
  {Schiemangk}}, \bibinfo {author} {\bibfnamefont {S.}~\bibnamefont
  {Spießberger}}, \bibinfo {author} {\bibfnamefont {A.}~\bibnamefont {Wicht}},
  \bibinfo {author} {\bibfnamefont {G.}~\bibnamefont {Erbert}}, \bibinfo
  {author} {\bibfnamefont {G.}~\bibnamefont {Tr\"ankle}},\ and\ \bibinfo
  {author} {\bibfnamefont {A.}~\bibnamefont {Peters}},\ }\href
  {https://doi.org/10.1364/AO.53.007138} {\bibfield  {journal} {\bibinfo
  {journal} {Applied Optics}\ }\textbf {\bibinfo {volume} {53}},\ \bibinfo
  {pages} {7138} (\bibinfo {year} {2014})}\BibitemShut {NoStop}%
\bibitem [{\citenamefont {Jang}\ \emph {et~al.}(2015)\citenamefont {Jang},
  \citenamefont {Erkintalo}, \citenamefont {Coen},\ and\ \citenamefont
  {Murdoch}}]{jang_temporal_2015}%
  \BibitemOpen
  \bibfield  {author} {\bibinfo {author} {\bibfnamefont {J.~K.}\ \bibnamefont
  {Jang}}, \bibinfo {author} {\bibfnamefont {M.}~\bibnamefont {Erkintalo}},
  \bibinfo {author} {\bibfnamefont {S.}~\bibnamefont {Coen}},\ and\ \bibinfo
  {author} {\bibfnamefont {S.~G.}\ \bibnamefont {Murdoch}},\ }\href
  {https://doi.org/10.1038/ncomms8370} {\bibfield  {journal} {\bibinfo
  {journal} {Nature Communications}\ }\textbf {\bibinfo {volume} {6}},\
  \bibinfo {pages} {7370} (\bibinfo {year} {2015})}\BibitemShut {NoStop}%
\bibitem [{\citenamefont {Hendry}\ \emph {et~al.}(2019)\citenamefont {Hendry},
  \citenamefont {Garbin}, \citenamefont {Murdoch}, \citenamefont {Coen},\ and\
  \citenamefont {Erkintalo}}]{hendry_impact_2019-1}%
  \BibitemOpen
  \bibfield  {author} {\bibinfo {author} {\bibfnamefont {I.}~\bibnamefont
  {Hendry}}, \bibinfo {author} {\bibfnamefont {B.}~\bibnamefont {Garbin}},
  \bibinfo {author} {\bibfnamefont {S.~G.}\ \bibnamefont {Murdoch}}, \bibinfo
  {author} {\bibfnamefont {S.}~\bibnamefont {Coen}},\ and\ \bibinfo {author}
  {\bibfnamefont {M.}~\bibnamefont {Erkintalo}},\ }\href
  {https://doi.org/10.1103/PhysRevA.100.023829} {\bibfield  {journal} {\bibinfo
   {journal} {Physical Review A}\ }\textbf {\bibinfo {volume} {100}},\ \bibinfo
  {pages} {023829} (\bibinfo {year} {2019})}\BibitemShut {NoStop}%
\bibitem [{\citenamefont {Javaloyes}\ \emph {et~al.}(2016)\citenamefont
  {Javaloyes}, \citenamefont {Camelin}, \citenamefont {Marconi},\ and\
  \citenamefont {Giudici}}]{javaloyes_dynamics_2016}%
  \BibitemOpen
  \bibfield  {author} {\bibinfo {author} {\bibfnamefont {J.}~\bibnamefont
  {Javaloyes}}, \bibinfo {author} {\bibfnamefont {P.}~\bibnamefont {Camelin}},
  \bibinfo {author} {\bibfnamefont {M.}~\bibnamefont {Marconi}},\ and\ \bibinfo
  {author} {\bibfnamefont {M.}~\bibnamefont {Giudici}},\ }\bibfield  {journal}
  {\bibinfo  {journal} {Physical Review Letters}\ }\textbf {\bibinfo {volume}
  {116}},\ \href {https://doi.org/10.1103/PhysRevLett.116.133901}
  {10.1103/PhysRevLett.116.133901} (\bibinfo {year} {2016})\BibitemShut
  {NoStop}%
\bibitem [{\citenamefont {Phillips}(1998)}]{phillips_nobel_1998}%
  \BibitemOpen
  \bibfield  {author} {\bibinfo {author} {\bibfnamefont {W.~D.}\ \bibnamefont
  {Phillips}},\ }\href {https://doi.org/10.1103/RevModPhys.70.721} {\bibfield
  {journal} {\bibinfo  {journal} {Reviews of Modern Physics}\ }\textbf
  {\bibinfo {volume} {70}},\ \bibinfo {pages} {721} (\bibinfo {year}
  {1998})}\BibitemShut {NoStop}%
\bibitem [{\citenamefont {Davenport}\ \emph {et~al.}(2018)\citenamefont
  {Davenport}, \citenamefont {Liu},\ and\ \citenamefont
  {Bowers}}]{davenport_integrated_2018}%
  \BibitemOpen
  \bibfield  {author} {\bibinfo {author} {\bibfnamefont {M.~L.}\ \bibnamefont
  {Davenport}}, \bibinfo {author} {\bibfnamefont {S.}~\bibnamefont {Liu}},\
  and\ \bibinfo {author} {\bibfnamefont {J.~E.}\ \bibnamefont {Bowers}},\
  }\href {https://doi.org/10.1364/PRJ.6.000468} {\bibfield  {journal} {\bibinfo
   {journal} {Photonics Research}\ }\textbf {\bibinfo {volume} {6}},\ \bibinfo
  {pages} {468} (\bibinfo {year} {2018})}\BibitemShut {NoStop}%
\bibitem [{\citenamefont {Delfyett}\ \emph {et~al.}(2019)\citenamefont
  {Delfyett}, \citenamefont {Klee}, \citenamefont {Bagnell}, \citenamefont
  {Juodawlkis}, \citenamefont {Plant},\ and\ \citenamefont
  {Zaman}}]{delfyett_exploring_2019}%
  \BibitemOpen
  \bibfield  {author} {\bibinfo {author} {\bibfnamefont {P.~J.}\ \bibnamefont
  {Delfyett}}, \bibinfo {author} {\bibfnamefont {A.}~\bibnamefont {Klee}},
  \bibinfo {author} {\bibfnamefont {K.}~\bibnamefont {Bagnell}}, \bibinfo
  {author} {\bibfnamefont {P.}~\bibnamefont {Juodawlkis}}, \bibinfo {author}
  {\bibfnamefont {J.}~\bibnamefont {Plant}},\ and\ \bibinfo {author}
  {\bibfnamefont {A.}~\bibnamefont {Zaman}},\ }\href
  {https://doi.org/10.1364/AO.58.000D39} {\bibfield  {journal} {\bibinfo
  {journal} {Applied Optics}\ }\textbf {\bibinfo {volume} {58}},\ \bibinfo
  {pages} {D39} (\bibinfo {year} {2019})}\BibitemShut {NoStop}%
\bibitem [{\citenamefont {Chen}\ \emph {et~al.}(2017)\citenamefont {Chen},
  \citenamefont {Ong}, \citenamefont {Ang}, \citenamefont {Lim}, \citenamefont
  {Png},\ and\ \citenamefont {Tan}}]{chen_broadband_2017}%
  \BibitemOpen
  \bibfield  {author} {\bibinfo {author} {\bibfnamefont {G.~F.~R.}\
  \bibnamefont {Chen}}, \bibinfo {author} {\bibfnamefont {J.~R.}\ \bibnamefont
  {Ong}}, \bibinfo {author} {\bibfnamefont {T.~Y.~L.}\ \bibnamefont {Ang}},
  \bibinfo {author} {\bibfnamefont {S.~T.}\ \bibnamefont {Lim}}, \bibinfo
  {author} {\bibfnamefont {C.~E.}\ \bibnamefont {Png}},\ and\ \bibinfo {author}
  {\bibfnamefont {D.~T.~H.}\ \bibnamefont {Tan}},\ }\bibfield  {journal}
  {\bibinfo  {journal} {Scientific Reports}\ }\textbf {\bibinfo {volume} {7}},\
  \href {https://doi.org/10.1038/s41598-017-07618-6}
  {10.1038/s41598-017-07618-6} (\bibinfo {year} {2017})\BibitemShut {NoStop}%
\bibitem [{\citenamefont {Moille}\ \emph {et~al.}(2019)\citenamefont {Moille},
  \citenamefont {Li}, \citenamefont {Briles}, \citenamefont {Yu}, \citenamefont
  {Drake}, \citenamefont {Lu}, \citenamefont {Rao}, \citenamefont {Westly},
  \citenamefont {Papp},\ and\ \citenamefont
  {Srinivasan}}]{moille_broadband_2019}%
  \BibitemOpen
  \bibfield  {author} {\bibinfo {author} {\bibfnamefont {G.}~\bibnamefont
  {Moille}}, \bibinfo {author} {\bibfnamefont {Q.}~\bibnamefont {Li}}, \bibinfo
  {author} {\bibfnamefont {T.~C.}\ \bibnamefont {Briles}}, \bibinfo {author}
  {\bibfnamefont {S.-P.}\ \bibnamefont {Yu}}, \bibinfo {author} {\bibfnamefont
  {T.}~\bibnamefont {Drake}}, \bibinfo {author} {\bibfnamefont
  {X.}~\bibnamefont {Lu}}, \bibinfo {author} {\bibfnamefont {A.}~\bibnamefont
  {Rao}}, \bibinfo {author} {\bibfnamefont {D.}~\bibnamefont {Westly}},
  \bibinfo {author} {\bibfnamefont {S.~B.}\ \bibnamefont {Papp}},\ and\
  \bibinfo {author} {\bibfnamefont {K.}~\bibnamefont {Srinivasan}},\ }\href
  {http://arxiv.org/abs/1907.11002} {\bibfield  {journal} {\bibinfo  {journal}
  {arXiv:1907.11002 [physics]}\ } (\bibinfo {year} {2019})},\ \bibinfo {note}
  {arXiv: 1907.11002}\BibitemShut {NoStop}%
\bibitem [{\citenamefont {Pfeiffer}\ \emph {et~al.}(2018)\citenamefont
  {Pfeiffer}, \citenamefont {Herkommer}, \citenamefont {Liu}, \citenamefont
  {Morais}, \citenamefont {Zervas}, \citenamefont {Geiselmann},\ and\
  \citenamefont {Kippenberg}}]{pfeiffer_photonic_2018}%
  \BibitemOpen
  \bibfield  {author} {\bibinfo {author} {\bibfnamefont {M.~H.~P.}\
  \bibnamefont {Pfeiffer}}, \bibinfo {author} {\bibfnamefont {C.}~\bibnamefont
  {Herkommer}}, \bibinfo {author} {\bibfnamefont {J.}~\bibnamefont {Liu}},
  \bibinfo {author} {\bibfnamefont {T.}~\bibnamefont {Morais}}, \bibinfo
  {author} {\bibfnamefont {M.}~\bibnamefont {Zervas}}, \bibinfo {author}
  {\bibfnamefont {M.}~\bibnamefont {Geiselmann}},\ and\ \bibinfo {author}
  {\bibfnamefont {T.~J.}\ \bibnamefont {Kippenberg}},\ }\href
  {https://doi.org/10.1109/JSTQE.2018.2808258} {\bibfield  {journal} {\bibinfo
  {journal} {IEEE Journal of Selected Topics in Quantum Electronics}\ }\textbf
  {\bibinfo {volume} {24}},\ \bibinfo {pages} {1} (\bibinfo {year}
  {2018})}\BibitemShut {NoStop}%
\bibitem [{\citenamefont {Liu}\ \emph {et~al.}(2016)\citenamefont {Liu},
  \citenamefont {Brasch}, \citenamefont {Pfeiffer}, \citenamefont {Kordts},
  \citenamefont {Kamel}, \citenamefont {Guo}, \citenamefont {Geiselmann},\ and\
  \citenamefont {Kippenberg}}]{liu_frequency-comb-assisted_2016}%
  \BibitemOpen
  \bibfield  {author} {\bibinfo {author} {\bibfnamefont {J.}~\bibnamefont
  {Liu}}, \bibinfo {author} {\bibfnamefont {V.}~\bibnamefont {Brasch}},
  \bibinfo {author} {\bibfnamefont {M.~H.~P.}\ \bibnamefont {Pfeiffer}},
  \bibinfo {author} {\bibfnamefont {A.}~\bibnamefont {Kordts}}, \bibinfo
  {author} {\bibfnamefont {A.~N.}\ \bibnamefont {Kamel}}, \bibinfo {author}
  {\bibfnamefont {H.}~\bibnamefont {Guo}}, \bibinfo {author} {\bibfnamefont
  {M.}~\bibnamefont {Geiselmann}},\ and\ \bibinfo {author} {\bibfnamefont
  {T.~J.}\ \bibnamefont {Kippenberg}},\ }\href
  {https://doi.org/10.1364/OL.41.003134} {\bibfield  {journal} {\bibinfo
  {journal} {Optics Letters}\ }\textbf {\bibinfo {volume} {41}},\ \bibinfo
  {pages} {3134} (\bibinfo {year} {2016})}\BibitemShut {NoStop}%
\end{thebibliography}%

\end{document}